\documentclass[aps, twocolumn, prd, preprintnumbers, amsmath, amssymb, amsfonts, superscriptaddress, nofootinbib]{revtex4-1}
\pdfoutput=1

\usepackage{amsmath,amssymb,bm,mathrsfs}
\usepackage{hyperref}
\usepackage{url}
\usepackage{graphicx}
\usepackage{color}
\usepackage{braket}
\usepackage{appendix}
\usepackage{comment}

\newcommand{\mr}[1]{\mathrm{#1}}
\newcommand{\mc}[1]{\mathcal{#1}}
\newcommand{\bs}[1]{\boldsymbol{#1}}

\definecolor{Orange}{cmyk}{0,0.61,0.87,0}
\definecolor{JungleGreen}{cmyk}{0.99,0,0.52,0}
\definecolor{OliveGreen}{cmyk}{0.64,0,0.95,0.40}
\definecolor{Brown}{cmyk}{0,0.81,1,0.60}
\definecolor{RoyalBlue}{cmyk}{0.71,0.53,0,0.12}
\definecolor{Gray}{cmyk}{0,0,0,0.40}
\definecolor{LightPink}{cmyk}{0.0,0.25,0,0}
\definecolor{LLightPink}{cmyk}{0.0,0.10,0,0}
\definecolor{LightBlue}{cmyk}{0.25,0,0,0}
\definecolor{LightGray}{cmyk}{0,0,0,0.2}

\begin{document}
\hfill TU-1303

\title{Hydrodynamics of Filtered Dark Matter: A Two-Component Approach
}

\author{Juntaro Wada}
\email{juntaro.wada.e5@tohoku.ac.jp }
\affiliation{Department of Physics, Tohoku University, Sendai, Miyagi 980-8578, Japan}
\affiliation{Technical University of Munich (TUM), School of Natural Sciences, Physics Department, James-Franck-Str. 1, 85748 Garching, Germany,
}

\begin{abstract}
We study the hydrodynamics of the Filtered Dark Matter (Filtered DM) scenario during a first-order phase transition (FOPT). In this scenario, the bubble wall is highly reflective of the dark matter (DM) fluid but transparent to radiation, making the hydrodynamic problem fundamentally different from that of the electroweak FOPT. Motivated by this property, we formulate the hydrodynamics of this system as a two-component fluid composed of DM and radiation, and find that the solutions can be classified into detonation-like and deflagration-like branches in the ballistic regime and in the local thermal equilibrium (LTE) regime. In the ballistic regime, the energy--momentum of DM that cannot enter the wall appears as a reflected mode, while in the LTE regime, it relaxes into the energy--momentum of radiation. We find that this difference in the fate of the DM fluid that cannot enter the interior of the wall leads to different hydrodynamic behaviors in the DM and radiation fluid independently and, in particular, results in different existence conditions for solutions in the deflagration-like branch. Based on these results, we further revisit the impact of hydrodynamic effects on the relic abundance of Filtered DM and demonstrate the change in the abundance induced by hydrodynamic effects.
In addition, we also discuss the non-conservation of the entropy current from the viewpoint of the two-fluid system,  and briefly comment on the similarity between the Filtered DM system and information-thermodynamic systems.
\end{abstract}

\maketitle

\section{Introduction}
Cosmological first-order phase transitions (FOPT) provide a well-motivated arena in which rich nonequilibrium phenomena in the early Universe can be realized. In fact, the out-of-equilibrium dynamics associated with expanding bubble walls can lead to a variety of cosmological consequences, including gravitational waves~\cite{Witten:1984rs, Hogan:1986dsh, Turner:1990rc, Kosowsky:1992vn, Kosowsky:1992rz, Kamionkowski:1993fg}, baryogenesis~\cite{Kuzmin:1985mm, Cohen:1993nk, Rubakov:1996vz}, and dark matter (DM) production~\cite{Falkowski:2012fb, Hambye:2018qjv, Baker:2019ndr, Chway:2019kft, Azatov:2021ifm, Azatov:2022tii, Jiang:2023nkj}. 

Among these phenomenologies associated with FOPT, the Filtered Dark Matter (Filtered DM) scenario provides a qualitatively different realization of DM production during a FOPT~\cite{Baker:2019ndr, Chway:2019kft}. (See also Refs.~~\cite{Chao:2020adk, Ahmadvand:2021vxs, Hosseini:2024ahe}) In this setup, DM is approximately in thermal equilibrium in the symmetric phase, while its mass becomes much larger in the broken phase. As a result, when a DM particle hits the bubble wall, only those particles whose momentum component normal to the wall is sufficiently large can penetrate into the broken phase, whereas the others are reflected or excluded by the wall~\cite{Baker:2019ndr, Chway:2019kft}. The final relic abundance is then controlled not directly by the conventional freeze-out process, but by the fraction of particles that successfully pass through the wall during FOPT, thereby allowing the scenario to evade the unitarity bound~\cite{Griest:1989wd}.

These properties strongly suggest that hydrodynamics should play a central role in the Filtered DM scenario. In particular, the temperature and velocity profiles around the wall affect the momentum distribution of incident dark matter particles and hence the filtering efficiency. Nevertheless, dedicated discussions of hydrodynamics in this context remain rather limited. In the original analysis of the filtering-out effect, Ref.~\cite{Chway:2019kft} incorporates hydrodynamic effects and, in particular, examines whether the Filtered DM abundance can reproduce the correct relic density for detonation solutions. A more dedicated study was later performed in Ref.~\cite{Jiang:2023nkj}, where the impact of hydrodynamic effects on the Filtered DM relic abundance was analyzed in detail.

We note that the hydrodynamic description appropriate for Filtered DM should differ conceptually from the conventional treatment that has been applied to the electroweak phase transition~\cite{Steinhardt:1981ct, Kurki-Suonio:1984zeb, Landau:1989, Laine:1993ey, Ignatius:1993qn, Bodeker:2009qy, Bodeker:2017cim, Hoche:2020ysm, Gouttenoire:2021kjv}. This is because, in the Filtered DM setup, the bubble wall distinguishes between DM and the rest of the radiation component with respect to their mass changes across the wall.  In addition, the relevant dynamics depend crucially on whether the DM component behaves ballistically near the wall or remains in local thermal equilibrium (LTE)~\cite{Ai:2024btx, Krajewski:2024xuz}. In the ballistic regime, the energy--momentum carried by DM particles that fail to enter the wall appears as a reflected component. In the LTE regime, by contrast, that energy--momentum is transferred to the radiation fluid through relaxation processes. However, these effects were not discussed in previous studies~\cite{Chway:2019kft, Jiang:2023nkj}.

In this work, motivated by these observations, we formulate a two-component hydrodynamic framework consisting of a DM fluid and a radiation fluid. We analyze the behavior of the DM fluid in the ballistic regime and the LTE regime, where a relatively model-independent treatment is possible, incorporating both the reflection effects at the wall and the energy--momentum transfer associated with relaxation processes. We also include the effect of the change in the sound speed across the wall~\cite{Leitao:2014pda, Giese:2020rtr, Giese:2020znk, Ai:2023see, Sanchez-Garitaonandia:2023zqz} and obtain analytic hydrodynamic solutions that allow a more robust determination of the impact on the Filtered DM relic abundance. Moreover, we revisit the entropy current, which has conventionally been considered to be conserved in the LTE regime~\cite{BarrosoMancha:2020fay, Balaji:2020yrx, Ai:2021kak}, and show that, from the viewpoint of the two-component system, it is not conserved even when LTE is maintained between the components, due to the conversion of energy--momentum from the DM sector to radiation.

This paper is organized as follows. In Sec.~\ref{sec: Filtered Dark Matter}, we review the Filtered DM scenario. In Sec.~\ref{sec: Two Component Hydrodynamics}, we formulate a hydrodynamic framework that distinguishes between the DM fluid and the radiation fluid, which we refer to as two-component hydrodynamics. Within this framework, we discuss the differences between the ballistic regime and the LTE regime, as well as the properties of the entropy current. In Sec.~\ref{sec: Matching condition on the wall}, we present the matching conditions derived from energy--momentum conservation across the wall based on the two-component hydrodynamics, and obtain their solutions analytically. In Sec.~\ref{sec: Impact of DM Relic Density}, using the results obtained in the previous sections, we demonstrate how the relic abundance of Filtered DM is affected by hydrodynamic effects. Finally, Sec.~\ref{sec: Summary} provides a summary of this paper.

\section{Filtered Dark Matter}
\label{sec: Filtered Dark Matter}
\subsection{Toy Model Setup}
To review the basic concepts for the Filtered DM scenario, we consider a simple concrete example; the dark-sector Lagrangian contains a fermionic dark matter particle $\chi$, as shown below~\cite{Baker:2019ndr, Chway:2019kft}:
\begin{align}
  \mc{L}&\supset - V(\phi) 
                 - y_\chi \phi \bar{\chi} \chi
                 - \beta \, \phi^2 H^\dagger H,
  \label{eq:model}
\end{align}
where $\phi$ is a dark scalar field that gives mass to $\chi$, and $H$ denotes the Standard Model (SM) Higgs field.  $y_\chi$ denotes the Yukawa coupling between the Filtered DM and the dark scalar, while $\beta$ represents the coupling between the dark scalar and the SM Higgs field, respectively.
We do not specify the explicit form of the dark scalar potential $V(\phi)$, but assume that $\phi$ acquires a vacuum expectation value $v_\phi := \braket{\phi}$, which induces dark matter mass $m_{\chi} : = y_{\chi} \braket{\phi}$. For simplicity, we assume that the mass of $\phi$ changes only negligibly across the phase transition. 
The stability of $\chi$ is assumed to be ensured by an appropriate symmetry, such as a $U(1)$ symmetry.

In the Filtered DM scenario, the FOPT of $\phi$ plays a crucial role in driving dark matter out of equilibrium. 
During the FOPT, $\chi$ is in thermal equilibrium outside the wall (the symmetric phase), while inside the wall (the broken phase), $\chi$ acquires a larger mass than outside the temperature. 
More quantitatively, this condition can be expressed as
$m_{\chi}^{\mr{in}} : = y_{\chi} v_{\phi} \gg T_{n}$,
where $T_{n}$ denotes the FOPT nucleation temperature. On the other hand, dark matter mass is dominantly determined by thermal mass outside the wall; $m_{\chi}^{\mr{out}} \simeq y_{\chi} T_{n}/4$.

\begin{figure}[t]
  \centering
  {\includegraphics[width=0.45\textwidth]{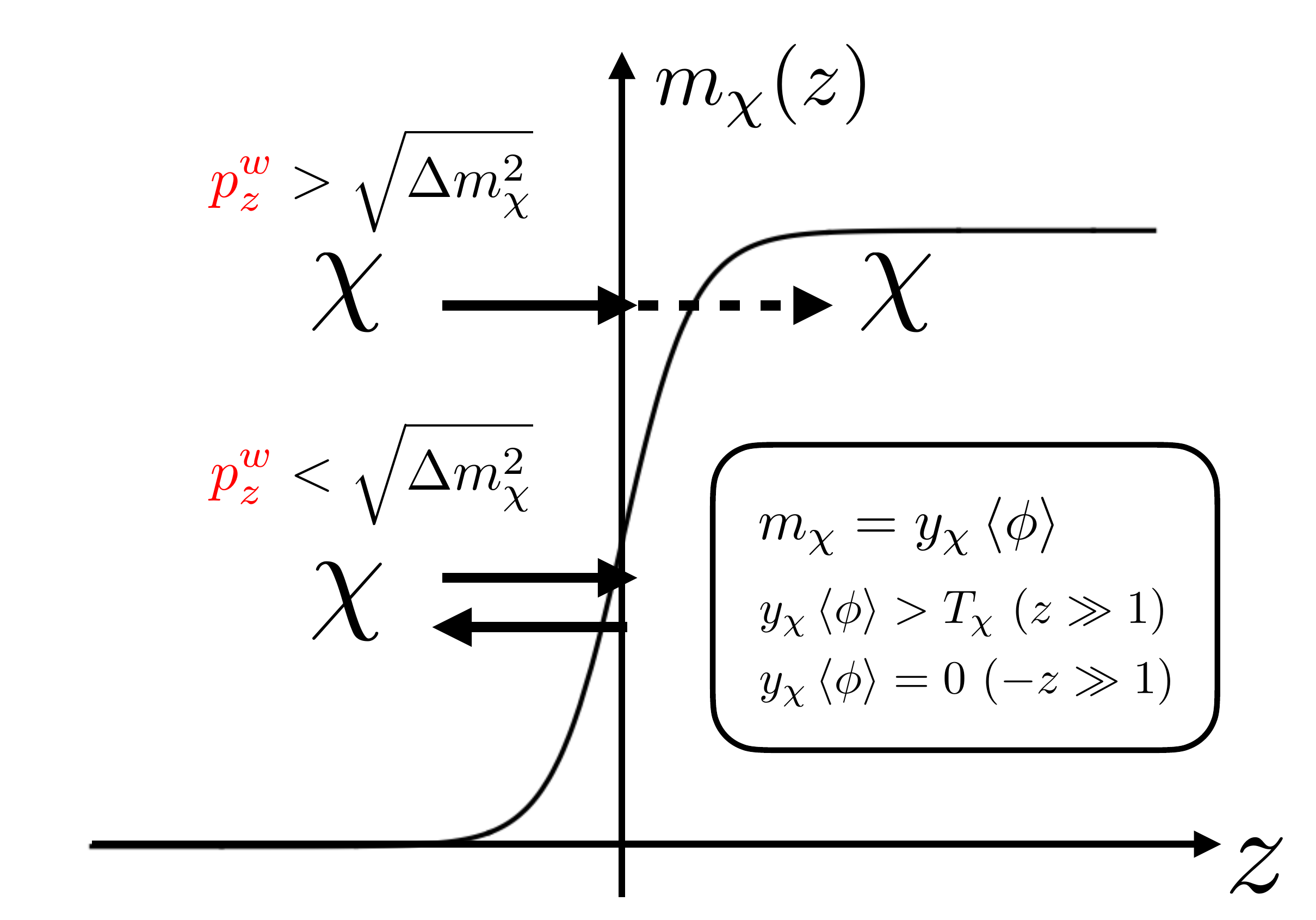}} 
  \caption{
      Basic concept of the filtered dark matter scenario.
  }
  \label{fig: Filtered DM}
\end{figure}

In this situation, energy conservation implies that only dark matter particles with sufficiently large momentum can penetrate into the interior of the wall upon collision, while the remaining particles are filtered out. In particular, if we adopt the planar-wall approximation and assume that the wall is propagating along the $z$ direction, this condition is given by~\cite{Baker:2019ndr, Chway:2019kft}
\begin{align}
    p_{z}^{w} > \sqrt{\Delta m_{\chi}^2},
\end{align}
where $p_{z}^{w}$ denotes the $z$ component of the dark matter momentum measured in the rest frame of the wall, and $\Delta m_{\chi}^2 = (m_{\chi}^{\mathrm{in}})^2 - (m_{\chi}^{\mathrm{out}})^2$. In Fig.~\ref{fig: Filtered DM}, we present a schematic overview of this setup.

\subsection{Filtering Eﬀect}
\label{subsec: Filtering Effect}
Hereafter, we set the wall position to be at $z = 0$ (see Fig.~\ref{fig: Filtered DM}). Outside the wall, the dark matter distribution function is taken to follow an equilibrium form. When boosted to the rest frame of the wall, it can be written as
\begin{align}
 f^{\mr{eq}}_{\chi}(z \ll 0, \bs{p}^w) 
	& \simeq 
	\exp\!\left[
		-\frac{\gamma_{w} E_{\chi}^{\mr{out},w} - v_{w} p_{z}^{w}}{T_{n}}
	\right],
\end{align}
where $E_{\chi}^{\mr{out},w} = \sqrt{|\bs{p}^w|^2 + (m_{\chi}^{\mr{out}})^2}$ is its energy measured in the wall rest frame. Hereafter, a subscript $w$ attached to the energy or momentum denotes the quantity evaluated in the wall rest frame. The wall velocity is denoted by $v_w$, with the associated Lorentz factor $\gamma_{w} = 1/\sqrt{1-v_w^2}$. Here and hereafter, the equilibrium distribution is approximated by the Maxwell-Boltzmann form.

The particle flux of dark matter incident on the wall from the false vacuum, per unit area and per unit time, is given by~\cite{Chway:2019kft, Jiang:2023nkj, Gehrman:2023qjn}
\begin{align}
\label{eq: DM flux}
J_{\chi}^{w}
	\simeq g_{\chi} \int \frac{d^3 p^w}{(2 \pi)^3} \frac{p_{z}^{w}}{|\bs{p}^w|} \,
	f^{\mr{eq}}_{\chi}\,
	\theta\!\left(p_{z}^w - \sqrt{\Delta m_{\chi}^2}\right),
\end{align}
where $g_\chi = 2$ denotes the internal degree of freedom of the Filtered DM.

The number density of dark matter particles that can penetrate into the interior of the wall,
$n_{\chi}^{\mr{in}} = J_{\chi}^w / (\gamma_{w} v_{w})$,
is obtained by explicitly evaluating Eq.~\eqref{eq: DM flux}, yielding~\cite{Chway:2019kft, Jiang:2023nkj, Gehrman:2023qjn}
\begin{align}
\label{eq: number density inside}
n_{\chi}^{\mr{in}} 
	\simeq
	\frac{g_{\chi} T_{n}^3}{\gamma_{w} v_w}
	\left(
	\frac{\gamma_{w}(1-v_w)m_{\chi}^{\mr{in}}/T_n + 1}{4 \pi^2 \gamma^3 (1-v_w)^2}
	\right)
	e^{- \gamma_w (1-v_w) m_{\chi}^{\mr{in}}/T_n } .
\end{align}

If the wall velocity is taken to be sufficiently small
$v_w = c \sqrt{T_n / m_{\chi}^{\mr{in}}} \ll 1$ and $\gamma_w \simeq 1$, with $c$ being an $\mc{O}(1)$ coefficient, we obtain
\begin{align}
n_{\chi}^{\mr{in}} 
	&\simeq 
	\frac{g_{\chi} (m_{\chi}^{\mr{in}} T_n)^{3/2}}{4 \pi^2 c}
	\left(1 + \frac{T_n}{m_{\chi}^{\mr{in}}}\right)
	e^{- m_{\chi}^{\mr{in}}/T_n} \nonumber \\
    \label{eq: DM number density NR without hydrodynamics}
	&\simeq 
	\frac{1}{4} \frac{(2 \pi)^{3/2}}{\pi^2 c}\,
	n_{\chi}^{\mr{in}, \mr{eq}}(T_n) \, .
\end{align}
Here $n_{\chi}^{\mr{in}, \mr{eq}}(T_n)$ is the equilibrium number density of DM inside the wall at the temperature $T_n$. In the first line we have used the approximation $\Delta m_{\chi} \simeq m_{\chi}^{\mr{in}}$.

Here, we expect that there should exist a cutoff in the velocity of dark matter particles for which the Filtered DM scenario works. This is because, in the limit where the wall is at rest, the right-hand side of \eqref{eq: DM number density NR without hydrodynamics} diverges. As one possible criterion, one may impose the condition that the wall velocity must be larger than the average speed of dark matter inside the wall:
\begin{align}
v_w \gtrsim v_{\chi}^{\mr{th}} ,
\end{align}
where $v_{\chi}^{\mr{th}} : = \sqrt{3 T_n / m_{\chi}^{\mr{in}}}$.
We then consider that if this condition is not satisfied, one needs to take into account the outflow of dark matter particles from inside the wall, induced by thermal fluctuations.
To neglect this effect, the parameter $c$ must satisfy the lower bound
\begin{align}
c \gtrsim \sqrt{3}.
\end{align}
For $c=\sqrt{3}$, we find $n_{\chi}^{\mr{in}} \simeq {1/4}\, n_{\chi}^{\mr{in}, \mr{eq}}$, which corresponds to the result obtained in Ref~\cite{Baker:2019ndr}. However, this criterion remains unsatisfactory from the viewpoint of hydrodynamics. In a later section~\ref{sec: Impact of DM Relic Density}, we will discuss a more plausible criterion.

In the relativistic limit, $v_w \simeq 1$ and $\gamma_w \to \infty$, one has $(1-v_w)\gamma_w \sim 1/(2\gamma_w)$. Therefore, the number density reduces to~\cite{Chway:2019kft, Jiang:2023nkj}
\begin{align}
n_{\chi}^{\mr{in}} \simeq g_{\chi}\frac{T_n^3}{\pi^2}
= n_{\chi}^{\mr{out},\mr{eq}}(T_n)\,,
\end{align}
where $n_{\chi}^{\mr{out}, \mr{eq}}(T_n)$ denotes the equilibrium number density of DM outside the wall at the temperature $T_n$.  
This implies that almost all DM particles outside the wall can penetrate into the wall if the wall velocity is sufficiently high.

\subsection{DM Relic Density (Without Hydrodynamics)}
When the wall velocity is low, the particle number density inside the wall is Boltzmann suppressed~\cite{Baker:2019ndr, Chway:2019kft}, and hence pair annihilation processes inside the wall are already decoupled. 
Moreover, under the assumption $m_{\chi}^{\mr{in}} \gg T_n$, pair production processes are also decoupled. 
Therefore, in this case, the DM number density freezes out at Eq.~\eqref{eq: DM number density NR without hydrodynamics}, and the present-day DM relic abundance is given by~\cite{Baker:2019ndr, Chway:2019kft}
\begin{align}
  \Omega_{\chi} h^2
    &\simeq 
    \frac{m_\chi \left(n_\chi^\mr{in} + n_{\bar{\chi}}^\mr{in}\right)}
         {3 M_{pl}^2 (H_0/h)^2}
    \frac{g_{* S 0} T_0^3}{g_{* S} T_n^3} \nonumber\\
    \label{eq: Filtered DM abundance}
    &\simeq 
    0.17 \, 
    \bigg( \frac{T_n}{\mr{TeV}} \bigg)
    \bigg( \frac{m_\chi^\mr{in} / T_n}{30} \bigg)^{5/2}
    \frac{e^{- m_\chi^\mr{in} / T_n}}{e^{-30}},
\end{align}
where $H_0 = 100\,h~\mathrm{km\,sec^{-1}\,Mpc^{-1}}$ is the Hubble constant with $h \simeq 0.67$~\cite{Planck:2018vyg}, and $T_0$ is the present temperature of the Universe. As is clear from this expression, in the case of a slow wall, the relic abundance is determined solely by the bubble dynamics and is independent of the detailed properties of the DM annihilation (and creation) channels in the model. This mechanism is referred to as the {\it Filtered DM scenario}~\cite{Baker:2019ndr, Chway:2019kft} and was proposed as a novel DM framework that evades the Griest-Kamionkowski bound~\cite{Griest:1989wd}.

On the other hand, when the wall velocity is relativistic, particles outside the wall penetrate into the interior with almost the same number density.
Inside the wall, pair production does not occur; however, the nonthermally populated DM particles continue to undergo pair annihilation until their number density is sufficiently depleted.
Since pair production has already decoupled inside the wall, the final DM relic abundance is not determined by the freeze-out temperature of the DM pair-production process, $T_{\mr{fo}}$, but instead by the lower nucleation temperature $T_n$.
Therefore, compared with the standard thermal relic abundance
$\left.\Omega_{\mr{DM}} h^2\right|_{\mr{thermal}}$~\cite{Ai:2024ikj},
\begin{align}
\label{eq: non thermal DM abundance}
  \Omega_{\mr{DM}} h^2
    = \frac{T_{\mr{fo}}}{T_n}
      \left.\Omega_{\mr{DM}} h^2\right|_{\mr{thermal}},
\end{align}
holds.
In other words, in this case, the relic abundance depends on the detailed interactions present in the model and is enhanced relative to the conventional thermal relic density.
This enhancement originates from the fact that the epoch at which the DM relic abundance is determined is delayed by the dynamics of the first-order phase transition.

Although this case (Eq.~\eqref{eq: non thermal DM abundance}) is not the main focus of our study, we note that in the two-component hydrodynamic system constructed later, when the relativistic limit of the wall speed is taken, the DM relic abundance is not estimated by Eq.~\eqref{eq: Filtered DM abundance} but rather by Eq.~\eqref{eq: non thermal DM abundance}.

\section{Two Component Hydrodynamics}
\label{sec: Two Component Hydrodynamics}
When discussing the hydrodynamics of Filtered DM, an important property is that only $\chi$ acquires a sizable mass difference between the inside and the outside of the wall. Namely,
\begin{equation}
\label{eq: derivative for mass profile in SM and dark sector}
\frac{d m_\chi^2}{d\phi} \neq 0, \qquad  \, \frac{d m_{a}^2}{d\phi} \simeq 0, (a \neq \chi),
\end{equation}
holds~\cite{Baker:2019ndr, Chway:2019kft}. Hereafter, particles that do not acquire mass across the wall are referred to as ``radiation''. Although particle excitations of the scalar field $\phi$ also acquire a mass due to the order parameter, we consider the case where this effect is negligible compared to the mass splitting of Filtered DM, and thus treat them as part of the radiation component. 

What is expected from Eq.~\eqref{eq: derivative for mass profile in SM and dark sector} is that the plasma fluid can no longer be treated as a single fluid; rather, its hydrodynamics should be described as a two-component system consisting of Filtered DM and a radiation component. Accordingly, we formulate the hydrodynamics of Filtered DM in terms of the \emph{two-component hydrodynamics}, which incorporates the effect that the Filtered DM fluid is reflected by the wall and the effect that its energy-momentum is converted into radiation through relaxation processes.

In this section, we develop a model-independent analysis, in the sense that we make use of Eq.~\eqref{eq: derivative for mass profile in SM and dark sector} but do not rely on any other model-dependent properties.

\subsection{Framework}
First, we decompose the whole energy-momentum tensor $T^{\mu\nu}$ into the scalar-field part $T^{\mu\nu}_\phi$, the radiation-fluid part $T^{\mu\nu}_{\mr{rad}}$, and the fluid part composed of Filtered DM, $T^{\mu\nu}_{\chi}$, as
\begin{equation}
T^{\mu\nu} = T^{\mu\nu}_\phi + T^{\mu\nu}_{\mr{rad}} + T^{\mu\nu}_{\chi} \, .
\end{equation}
These tensors satisfy the total energy-momentum conservation,
\begin{equation}
\label{eq: energy momentum conservation for whole system}
\nabla_{\mu} T^{\mu\nu} = 0\,.
\end{equation}
Hereafter, we assume that the time scale of bubble expansion is much shorter than the Hubble time. Under this assumption, the covariant derivative can be approximated by the partial derivative, i.e., $\nabla_\mu \simeq \partial_\mu$.

The evolution equations for each component of the fluid generically take the following form:
\begin{align}
\label{eq: energy momentum evolution for dark matter plasma}
\partial_{\mu} T_{\chi}^{\mu\nu} 
	&= \partial^{\nu} \phi \, \frac{d m_{\chi}^2}{d\phi} \int \frac{d^3 \bs{p}}{(2\pi)^3} \frac{1}{2 E_{\chi}} f_{\chi} 
		+ \Theta_{\chi}^{\nu}, \\
\label{eq: energy momentum evolution for radiation plasma}
\partial_{\mu} T_{\mr{rad}}^{\mu\nu} 
	&= \Theta_{\mr{rad}}^{\nu}.
\end{align}
Here, $f_{\chi}$ denotes the distribution function of the Filtered DM, and 
$E_{\chi} := \sqrt{m_{\chi}^2 + \bs{p}^2}$ is the energy corresponding to the momentum $\bs{p}$. 
The terms $\Theta_{\mr{rad}}^{\nu}$ and $\Theta_{\chi}^{\nu}$ represent the exchange of energy and momentum between different fluids, as well as between the fluids and the scalar-field wall.
From the viewpoint of the Boltzmann equation (More explicitly, Kadanoff-Baym equations), the microscopic origin of $\Theta_{\mr{rad}}^{\nu}$ and $\Theta_{\chi}^{\nu}$ can be understood as the collision terms induced by the particle interactions~\cite{Ramsey-Musolf:2025jyk, Ai:2025bjw}.\footnote{
More explicitly, when the exchange of energy and momentum is induced by the microscopic particle interactions, 
$\Theta_{\mr{rad}}^{\nu}$ and $\Theta_{\chi}^{\nu}$ can be expressed as
\begin{align}
\Theta_{\mr{rad}}^{\nu}
	 &= \sum_{a \in \mr{rad}} \int \frac{d^3 \bs{p}}{(2\pi)^3} p^{\nu} \, C[f_a], \\
\Theta_{\chi}^{\nu}
	&=  \int \frac{d^3 \bs{p}}{(2\pi)^3} p^{\nu} \, C[f_\chi].
\end{align}
Here, we denote the scattering (collision) term by
\begin{align}
\label{eq: definition for collision term}
C[f_a] &= - f_a(x, p_a) \prod_{i \neq a} n_i \int 
\frac{ \braket{ f | i \mathcal{T} | i, a } \braket{ i, a | i \mathcal{T} | f } }
{ T \prod_{i} (2 E_i)\, V } \nonumber \\
&\quad
\prod_f \frac{d^3 \vec{p}_f\, [1 \pm f_f(x, p_f)]}{2 E_f (2\pi)^3} 
+ \text{inverse process},
\end{align}
where $\braket{ i, a | i \mathcal{T} | f }$ denotes the matrix element for a process with an initial state that includes the particle species $a$, and $n_i$ is the number density of the incoming species $i$.
}

From the total energy--momentum conservation~\eqref{eq: energy momentum conservation for whole system}, the evolution of the energy--momentum tensor of the scalar field can be written as
\begin{align}
\label{eq: energy momentum evolution for scalar field}
\partial_{\mu} T_{\phi}^{\mu\nu} 
	&= - \partial^{\nu} \phi \, \frac{d m_{\chi}^2}{d\phi} 
	\int \frac{d^3 \bs{p}}{(2\pi)^3} \frac{1}{2 E_{\chi}} f_{\chi} 
	- \left( \Theta_{\chi}^{\nu} + \Theta_{\mr{rad}}^{\nu} \right) .
\end{align}
In the following, for simplicity, we assume
\begin{align}
\label{eq: energy-momentum balance}
 \Theta_{\chi}^{\nu} + \Theta_{\mr{rad}}^{\nu} \simeq 0 .
\end{align}
This assumption corresponds to a situation in which the exchange of energy and momentum between the fluid and the wall can be neglected.\footnote{This assumption breaks down when the wall velocity approaches the speed of light. In this regime, the friction acting on the wall is predominantly generated by backreaction from particle splitting processes~\cite{Bodeker:2009qy, Bodeker:2017cim, Hoche:2020ysm, Gouttenoire:2021kjv}.
}

Since the energy-momentum tensor of the scalar field is given by
\begin{align}
\label{eq: energy momentum tensor for scalar field}
T^{\mu\nu}_\phi 
&= (\partial^\mu \phi)(\partial^\nu \phi) 
 - g^{\mu\nu} \left( \frac{1}{2}(\partial\phi)^2 - V_0(\phi) \right) \,,
\end{align}
with $V_0(\phi)$ denoting the zero-temperature scalar potential, one can derive the equation of motion for the scalar field from
Eq.~\eqref{eq: energy momentum evolution for scalar field} as~\cite{Moore:1995ua, Moore:1995si}
\begin{align}
\label{eq: scalar eom}
\Box \phi 
+ \frac{\partial V_0}{\partial \phi} 
+ \frac{d m_{\chi}^2}{d\phi} 
	\int \frac{d^3 \bs{p}}{(2\pi)^3} \frac{1}{2 E_{\chi}} f_{\chi}
= 0 \,.
\end{align}

Once the distribution functions are specified, the explicit forms of 
$T_{\mr{rad}}^{\mu\nu}$ and $T_{\chi}^{\mu\nu}$ are given by
\begin{align}
T^{\mu\nu}_{\chi} 
&= \int {d^3 p \over(2 \pi)^3  E_{\chi}} \, p^{\mu} p^{\nu} \, f_{\chi}, \\
T^{\mu\nu}_{\mr{rad}} 
&= \sum_{a \in \mr{rad}} \int {d^3 p \over (2 \pi)^3 E_{a}} \, p^{\mu} p^{\nu} \, f_{a}.
\end{align}

We adopt the  LTE approximation~\cite{BarrosoMancha:2020fay, Balaji:2020yrx, Ai:2021kak} for radiation
species.
Namely, each radiation distribution function is decomposed into the locally
equilibrium part and a perturbative deviation from it,
\begin{align}
f_a &= f_a^{\mr{eq}} + \delta f_a \\
    &\simeq f_a^{\mr{eq}},
\end{align}
and we assume that $\delta f_a$  is sufficiently small to be neglected.\footnote{We recall that, as the equilibrium distribution function, we adopt the Boltzmann distribution $e^{-p \cdot u_{\mr{rad}}/T_{\mr{rad}}}$. Here, $u_{\mr{rad}}^{\mu}$ and $T_{\mr{rad}}$ denote the velocity and temperature of the radiation fluid, respectively.}

For the Filtered DM fluid, we decompose the distribution function into an
incident component toward the wall, $f_{\chi,f}$, and a reflected component $f_{\chi,r}$. For the incident component, we decomposed it into the locally equilibrium part and a perturbative deviation from it:
\begin{align}
&f_{\chi} = f_{\chi,f} + f_{\chi,r} , 
\\
&f_{\chi,f} = f_{\chi}^{\mr{eq}} + \delta f_{\chi,f},
\end{align}
and we regards $\delta f_{\chi,f}$ as small perturbation to be neglected hereafter.

For later convenience, we further decompose the energy--momentum tensor of the
DM fluid into the incident and reflected contributions as
\begin{align}
T_{\chi}^{\mu\nu} = T_{\chi,f}^{\mu\nu} + T_{\chi,r}^{\mu\nu} ,
\end{align}
and define
\begin{align}
\label{eq: def xi}
\Xi_{\chi}^{\nu} = \partial_{\mu} T_{\chi,r}^{\mu\nu} .
\end{align}

In the limit where the distinction between DM and radiation can be neglected, the effects of reflection and energy--momentum transfer between the fluids become irrelevant. In this case, the above framework reduces to the well-known hydrodynamics of the wall--plasma system~\cite{Steinhardt:1981ct, Kurki-Suonio:1984zeb, Landau:1989, Laine:1993ey, Ignatius:1993qn}. We refer to this case as \emph{single-component hydrodynamics} to distinguish it from our framework.

\subsection{Ballistic vs LTE}
The hydrodynamics during a FOPT is governed by three length scales: the bubble radius $R_{\text{bubble}}$, the mean free path in the plasma frame $L_{\text{MFP}}$, and the wall width in the wall rest frame $L_w$~\cite{Ai:2024btx, Krajewski:2024xuz}. We focus on the regime $R_{\text{bubble}} \gg L_{\text{MFP}},\, L_w/\gamma_w$, where the system is macroscopic, and the fluid temperature and velocity are well defined, though not necessarily unchanged across it. As argued in Ref.~\cite{Jiang:2023nkj}, such a change in fluid temperature and velocity can affect the filtering effect and the Filtered DM relic abundance. Depending on the hierarchy between the remaining two scales, the fluid hydrodynamics falls into two distinct regimes~\cite{Ai:2024btx, Krajewski:2024xuz}:
\begin{itemize}
    \item $R_{\text{bubble}} \gg  L_{\text{MFP}} \gg L_w/\gamma_w \quad (\text{ballistic condition})$
    \item $R_{\text{bubble}} \gg L_w/\gamma_w \gg L_{\text{MFP}} \quad (\text{LTE condition})$
\end{itemize}

First, we consider the ballistic regime, in which the interaction rate $\Gamma \propto 1/L_{\text{MFP}}$ is very small. On the scale comparable to the wall thickness, the Filtered DM fluid and radiation are decoupled and behave as separate fluids, although they are in thermal equilibrium outside the wall on the bubble scale $R_{\text{bubble}}$. Consequently, near the wall, the energy--momentum exchange between the fluids can be neglected, and the Filtered DM components that do not enter the wall effectively behave as a reflected mode. In summary, reflection dominates in the ballistic regime, while inter-fluid energy--momentum exchange is negligible:
\begin{align}
|\Xi_{\chi}^{\nu}| \gg |\Theta_{\chi}^{\nu}|.\quad (\text{ballistic regime})
\end{align}

Next, we consider the LTE regime. In this case, the interaction rate $\Gamma \propto 1/L_{\text{MFP}}$ is sufficiently large even on the scale of the wall, so that local thermal equilibrium is maintained and reflection effects at the wall are rapidly relaxed within the fluid~\cite{Ai:2024btx, Krajewski:2024xuz}. As a result, reflected modes can be neglected. Instead, the energy--momentum of the DM fluid components that fail to enter the interior of the wall must be converted into radiation energy--momentum through relaxation processes.\footnote{
If $\Theta_{\chi}^{\nu}$ originates from particle interactions, one should not assume LTE for the distribution function when evaluating Eq.~\eqref{eq: definition for collision term}. This is because energy--momentum transport occurs through relaxation, i.e.\ the process by which $f_{\chi}$ returns to the LTE distribution. The LTE distribution used to compute the fluid energy--momentum is valid only on macroscopic scales of order $L_w/\gamma_w$, while on microscopic scales of order $L_{\text{MFP}}$ the system undergoes frequent relaxation.} In summary, in the LTE regime, the exchange of energy--momentum between the fluids is essential, while reflection effects can be neglected:
\begin{align}
|\Xi_{\chi}^{\nu}| \ll |\Theta_{\chi}^{\nu}|.\quad (\text{LTE regime})
\end{align}

In the next section, we solve the energy--momentum conservation equations across the wall in both the ballistic and LTE regimes to understand hydrodynamical effects. Interestingly, in both cases—where reflection dominates in the ballistic regime and where inter-fluid energy--momentum exchange is crucial in the LTE regime—the solutions can be classified into detonation-like and deflagration-like branches. 

\subsection{Comments on Entropy current}
In discussions of hydrodynamics during an FOPT, one often introduces, in addition to the energy--momentum tensor, a quantity known as the entropy current,
\begin{align}
S^{\mu}_{\chi, f}
	:= s_{\chi, f} \, u^{\mu}_{\chi, f} , \\
S^{\mu}_{\mr{rad}}
	:= s_{\mr{rad}} \, u^{\mu}_{\mr{rad}} .
\end{align}
Here, $s_{\chi, f(\mr{rad})}$ denotes the entropy density and $u^{\mu}_{\chi, f(\mr{rad})}$ the fluid four-velocity. The subscripts $(\chi, f)$ and $(\mr{rad})$ indicate that the quantity refers to the incident dark matter fluid or to the radiation fluid, respectively.

In single-component hydrodynamics, it is well known that the entropy current is conserved in the LTE regime~\cite{BarrosoMancha:2020fay, Balaji:2020yrx, Ai:2021kak}. In contrast, in two-component hydrodynamics, one can show that the entropy current is not conserved, even in the LTE regime, as we will show.

First, the entropy currents associated with each fluid can be written in the following form:\footnote{
From the viewpoint of the Boltzmann equation, the number flux $n_{\chi,f} u_{\chi,f}^{\mu}$ can be written as
\begin{align}
\partial_{\mu} (n_{\chi,f} u_{\chi,f}^{\mu})
= \int \frac{d^3 \bs{p}}{(2\pi)^3}  \, C[f_\chi] .
\end{align}
Therefore, assuming that the origin of $\Theta_{\chi}^{\nu}$ is given by the scattering term, the right-hand side of Eq.~\eqref{eq: entropy non-conservation in dark sector} can be summarized in the LTE regime as
\begin{align}
\partial_{\mu} S_{\chi,f}^{\mu}
	&= {1 \over T_{\chi,f}} \int \frac{d^3 \bs{p}}{(2\pi)^3}
	   (u_{\chi,f}^{\mu} p_{\mu}  - \mu_{\chi,f} )\, C[f_\chi] .
\end{align}
When the wall is not relativistic, one has
$u_{\chi,f}^{\mu} p_{\mu} \simeq p^0 \gtrsim \mu_{\chi,f} = m_{\chi,f}$.
This means that the entropy current for DM fluid decreases if reflection is neglected,
\begin{align}
\partial_{\mu} S_{\chi,f}^{\mu} < 0.
\end{align}
On the other hand, the entropy current increases in the radiation sector,
\begin{align}
\partial_{\mu} S_{\mr{rad}}^{\mu} > 0.
\end{align}
}
\begin{align}
\label{eq: entropy non-conservation in dark sector}
\partial_{\mu} S_{\chi, f}^{\mu}
	 &=-{\mu_{\chi} \over T_{\chi, f}} \, \partial_{\mu} \bigl(n_{\chi, f} \, u_{\chi, f}^{\mu}\bigr)
	   + {u_{\chi, f}^{\mu} \over T_{\chi,f}} \, \left(\Theta_{\chi, \mu} - \Xi_{\chi, \mu} \right)\nonumber\\
    &\quad
        +{u_{\chi,f}^{\mu} \over T_{\chi,f}}
            (\partial_{\mu} \mu_{\chi})
            \int \frac{d^3 p}{(2\pi)^3} {1 \over E_{\chi}} f_{\chi, r} ,
\\
\label{eq: entropy non-conservation in SM sector}
\partial_{\mu} S_{\mr{rad}}^{\mu} 
    &=-{\mu_{\mr{rad}} \over T_{\mr{rad}}} \, \partial_{\mu} \bigl(n_{\mr{rad}} \, u_{\mr{rad}}^{\mu}\bigr)
      + {u_{\mr{rad}}^{\mu} \over T_{\mr{rad}}} \, \Theta_{\mr{rad}, \mu}. 
\end{align}
Here, $\mu_{\chi(\mr{rad})}$ and $n_{\chi, f(\mr{rad})}$ denote the chemical potential and the number density, respectively.\footnote{
We note that the entropy density $s$ can be expressed as
\begin{align}
s
	= {\rho + p - \mu n \over T},
\end{align}
where $\rho$ and $p$ denote the energy density and pressure, respectively. For simplicity, the subscripts have been omitted.
} 
For the dark matter fluid, one has $\mu_{\chi} = m_{\chi}$ if LTE remains for the incident DM component, while for the radiation fluid $\mu_{\mr{rad}} \simeq 0$. We present the derivations of Eqs.~\eqref{eq: entropy non-conservation in dark sector} and \eqref{eq: entropy non-conservation in SM sector} in Appendix~\ref{app: Derivation of entropy current}.

Defining the total entropy current of the fluid as the sum of the two entropy currents,
$S^{\mu} = S^{\mu}_{\chi,f} + S^{\mu}_{\mr{rad}}$, it satisfies the following evolution equation:
\begin{align}
\label{eq: entropy non-conservation}
\partial_{\mu} S^{\mu}
	&\simeq 
        -{\mu_{\chi, f} \over T_{\chi, f}} \partial_{\mu} (n_{\chi, f}          u_{\chi, f}^{\mu})\nonumber\\
    &+
        \left({u_{\chi, f}^{\mu} \over T_{\chi,f}} - {u_{\mr{rad}}^{\mu} \over T_{\mr{rad}}}  \right)\Theta_{\chi}
         - {u_{\chi, f}^{\mu} \over T_{\chi,f}} \, 
         \Xi_{\chi, \mu} \nonumber\\
    &+
        {u_{\chi,f}^{\mu} \over T_{\chi,f}}
            (\partial_{\mu} \mu_{\chi})
            \int \frac{d^3 \bs{p}}{(2\pi)^3} {1 \over E_{\chi}}  f_{\chi, r} .
\end{align}

Therefore, both reflection and the energy--momentum exchange between the fluids affect the evolution of the entropy current. Even if one considers the LTE regime where reflection can be neglected, Eq.~\eqref{eq: entropy non-conservation} reduces to
\begin{align}
\label{eq: entropy non-conservation in LTE}
\partial_{\mu} S^{\mu}
	&\simeq 
        -{\mu_{\chi} \over T_{\chi, f}} \partial_{\mu} (n_{\chi, f}          u_{\chi, f}^{\mu})+
		\left({u_{\chi, f}^{\mu} \over T_{\chi,f}} - {u_{\mr{rad}}^{\mu} \over T_{\mr{rad}}}  \right)\Theta_{\chi},
\end{align}
which still implies non-conservation.

From the next section, we discuss the matching conditions for the fluid temperature and velocity across the wall that follow from energy--momentum conservation. If the entropy current were conserved, its conservation would provide an additional independent matching condition~\cite{BarrosoMancha:2020fay, Balaji:2020yrx, Ai:2021kak}. However, the non-conservation in Eq.~\eqref{eq: entropy non-conservation} makes it difficult to construct a closed set of matching conditions based on the evolution of the entropy current. Therefore, we do not pursue this issue further in this work.

\section{Matching condition on the wall }
\label{sec: Matching condition on the wall}
The hydrodynamics of the fluid can be understood by solving energy-momentum conservation in the vicinity of the wall~\cite{Steinhardt:1981ct, Kurki-Suonio:1984zeb, Landau:1989, Laine:1993ey, Ignatius:1993qn}. To see this, we work in the rest frame of the wall and adopt the planar approximation, taking the $z$-direction to be aligned with the wall expansion.

In this setup, among the energy--momentum conservation equations for the Filtered DM fluid and the radiation plasma, Eqs.~\eqref{eq: energy momentum evolution for dark matter plasma} and \eqref{eq: energy momentum evolution for radiation plasma}, the only nontrivial equations are $0$ and $z$ components:
\begin{align}
\partial_{z} T_{\mr{rad}(\chi)}^{z0}(z), \quad \partial_{z} T_{\mr{rad}(\chi)}^{zz}(z).
\end{align}
Moreover, the four-velocities of the radiation and dark matter fluids can be parametrized as
$u_{\mr{rad}}^{\mu}(z) \simeq \gamma_{\mr{rad}}(z)\,(1,0,0,v_{\mr{rad}}(z))$
and
$u_{\chi,f}^{\mu}(z) \simeq \gamma_{\chi,f}(z)\,(1,0,0,v_{\chi,f}(z))$, with $\gamma_{\chi,f(\mr{rad})} = 1/\sqrt{1-v_{\chi,f(\mr{rad})}^2}$ denoting the Lorenz factor, respectively.

\begin{figure}[t]
  \centering
  {\includegraphics[width=0.45\textwidth]{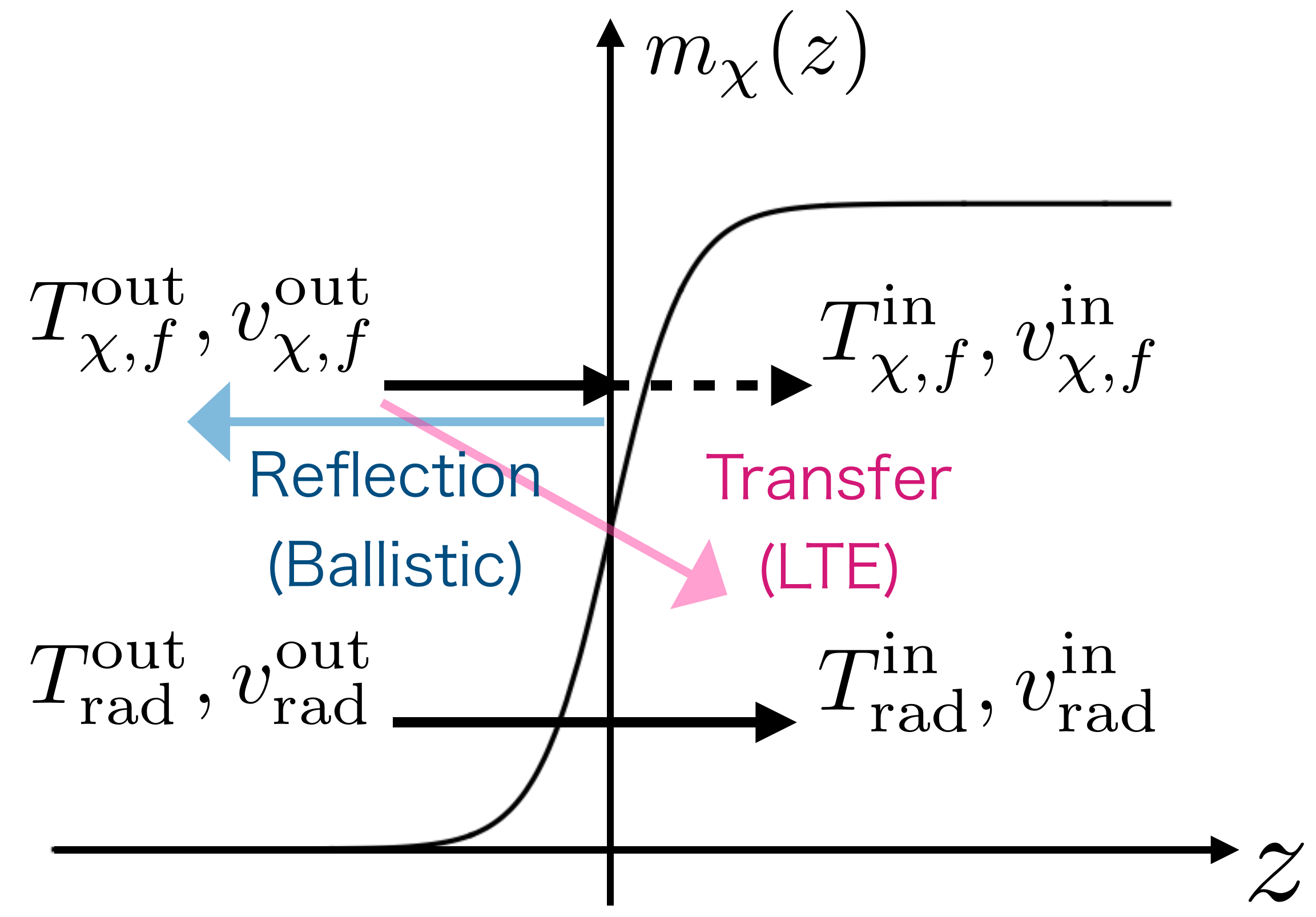}} 
  \caption{
      Schematic illustration of the variables: the temperatures and velocities inside and outside the wall. The subscripts $(\chi, f)$ and $(\mr{rad})$ label the quantities associated with the Filtered DM fluid and the radiation fluid. In the ballistic regime, hydrodynamics is governed by energy--momentum reflection, while in the LTE regime, it is governed by energy--momentum transfer from the Filtered DM fluid to the radiation fluid.
  }
  \label{fig: Two-component hydrodynamics}
\end{figure}

We note that, under the matching condition, the hydrodynamics is characterized by the temperatures and velocities of each fluid inside and outside the wall~\cite{Leitao:2014pda, Giese:2020rtr, Giese:2020znk, Ai:2023see, Sanchez-Garitaonandia:2023zqz}:
\begin{align}
T_{\chi,f(\mr{rad})}^{\mr{out}(\mr{in})}, \, v_{\chi,f(\mr{rad})}^{\mr{out}(\mr{in})}.
\end{align}
Here, the subscripts $(\mr{in})$ and $(\mr{out})$ denote quantities evaluated inside and outside the wall, while $(\chi,f)$ and $(\mr{rad})$ label the incident dark matter fluid and the radiation fluid, respectively. Figure~\ref{fig: Two-component hydrodynamics} shows a schematic configuration of these variables. The energy--momentum conservation equations provide the matching conditions among them.

We also note that outside the wall, the Filtered DM and radiation components are in thermal equilibrium. Thus, we can impose the following boundary conditions:
\begin{align}
\label{eq: assumption outside wall temperature}
&T_{\chi, f}^{\mr{out}} = T_{\mr{rad}}^{\mr{out}},\\
\label{eq: assumption outside wall velocity}
&v_{\chi,f}^{\mr{out}} = v_{\mr{rad}}^{\mr{out}}.
\end{align}

\subsection{Ballistic Regime}
We first consider the ballistic regime, in which the inter-fluid energy--momentum exchange can be neglected. In this case, since the Filtered DM fluid and the radiation are decoupled in the vicinity of the wall, it suffices to solve the energy--momentum conservation equations for the Filtered DM sector alone. Integrating both sides of Eq.~\eqref{eq: energy momentum evolution for dark matter plasma} over $z$, we obtain the following two matching conditions:
\begin{align}
&\omega_{\chi, f}^{\mr{out}} (\gamma_{\chi, f}^{\mr{out}})^2 v_{\chi, f}^{\mr{out}} 
- \omega_{\chi, f}^{\mr{in}} (\gamma_{\chi, f}^{\mr{in}})^2 v_{\chi, f}^{\mr{in}} \nonumber \\
	&\hspace{9em}
        =r \, \omega_{\chi, f}^{\mr{out}} (\gamma_{\chi, f}^{\mr{out}})^2 v_{\chi, f}^{\mr{out}},  
        \label{eq: Ballistic matching for energy flux in dark sector}\\[0.8em]
&\omega_{\chi, f}^{\mr{out}} (\gamma_{\chi, f}^{\mr{out}})^2 (v_{\chi, f}^{\mr{out}})^2 
	- \omega_{\chi, f}^{\mr{in}} (\gamma_{\chi, f}^{\mr{in}})^2 (v_{\chi, f}^{\mr{in}})^2 \nonumber \\
    &\hspace{2em}
		=   p_{\chi,f }^{\mr{in}} - (1 - r ) p_{\chi,f}^{\mr{out}} 
                + \Delta \epsilon_{\chi} + \Delta_{r}
        \nonumber \\
    &\hspace{5em}
			+ r \, \omega_{\chi, f}^{\mr{out}} (\gamma_{\chi, f}^{\mr{out}})^2 (v_{\chi, f}^{\mr{out}})^2,  
        \label{eq: Ballistic matching for momentum in dark sector}
\end{align}
where
\begin{align}
\omega_{\chi, f} := \rho_{\chi, f} + p_{\chi, f},  
\end{align}
denotes the enthalpy density of the DM fluid, while $\rho_{\chi, f}$ and $p_{\chi, f}$ are the corresponding energy density and pressure, respectively. $\Delta \epsilon_{\chi}$ denotes the latent heat.

Outside the wall, the Filtered DM behaves as a relativistic particle, and thus the energy density and pressure are given by
\begin{align}
\rho_{\chi,f}^{\mr{out}}
	&= g_{\chi}{\pi^2 \over 30} \left(T_{\chi,f}^{\mr{out}}\right)^4, \\
p_{\chi,f}^{\mr{out}}
	&= {1 \over 3}\rho_{\chi,f}^{\mr{out}},
\end{align}
where we have neglected the numerical factor associated with the difference between bosons and fermions.

The parameters $r$ and $\Delta_r$ characterize the fraction of the energy flux and momentum of the Filtered DM fluid that is reflected at the wall. They are defined as
\begin{align}
r
    &:= {\int dz \,\Xi_{\chi}^0 \over \omega_{\chi, r}^{\mr{out}} (\gamma_{\chi, f}^{\mr{out}})^2 v_{\chi, f}^{\mr{out}} } >0,  \\
\label{eq: anzatz for reflection}
\Delta_r
    &:=
    \int dz \,\Xi_{\chi}^z 
    - r \left(
        \,\omega_{\chi, r}^{\mr{out}} (\gamma_{\chi, f}^{\mr{out}})^2 (v_{\chi, f}^{\mr{out}})^2 + p_{\chi, f}^{\mr{out}}
        \right) < 0.
\end{align}

The parameter $r$, which represents the reflection coefficient of the energy flux, can be evaluated using the same method as that employed to compute the flux in Sec.~\ref{subsec: Filtering Effect}. Namely,
\begin{align}
\label{eq: evaluation of reflection rate}
r 
    &\simeq 1- {F_w \over \omega_{\chi, f}^{\mr{out}} (\gamma_{\chi, f}^{\mr{out}})^2 v_{\chi, f}^{\mr{out}}}, 
\end{align}
where
\begin{align}
F_w
	&:= g_{\chi} \int \frac{d^3 p^w}{(2 \pi)^3} p_{z}^{w} \,
	f^{\mr{eq}}_{\chi}\,
	\theta\!\left(p_{z}^w - \sqrt{\Delta m_{\chi}^2}\right),\\
    &={g_{\chi} \over 4 \pi^2}
        {(T_{\chi,f}^{\mr{out}})^4 \over (1-v_{\chi,f}^{\mr{out}})^3
        (\gamma_{\chi,f}^{\mr{out}})^4} \nonumber \\
    &\qquad
        \Big(3-v_{\chi,f}^{\mr{out}}
            + m_{\chi}^{\mr{in}} 
            (3-v_{\chi,f}^{\mr{out}})
            (1-v_{\chi,f}^{\mr{out}})\gamma_{\chi,f}^{\mr{out}}/T_{\chi,f}^{\mr{out}} \nonumber\\
    &\qquad
        +(m_{\chi}^{\mr{out}})^2 (1-v_{\chi,f}^{\mr{out}})^{2} (\gamma_{\chi,f}^{\mr{out}})^2/(T_{\chi,f}^{\mr{out}})^2
            \Big)\nonumber\\
    &\qquad
       \times e^{- \gamma_{\chi,f}^{\mr{out}} (1-v_{\chi,f}^{\mr{out}}) m_{\chi}^{\mr{in}}/T_{\chi,f}^{\mr{out}} }.
\end{align}

\begin{figure}[t]
  \centering
  {\includegraphics[width=0.47\textwidth]{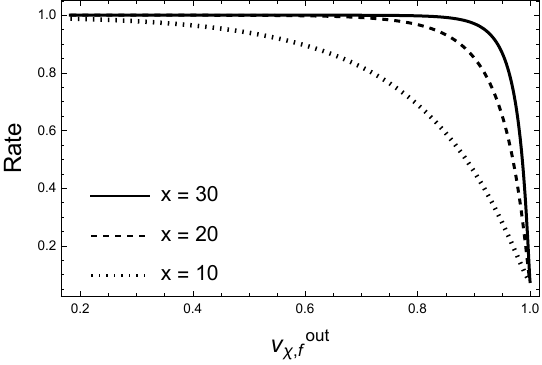}} 
  \caption{
     Velocity dependence of the coefficient characterizing the DM energy flux, interpreted as the reflection coefficient $r$ in the ballistic regime and as the transfer coefficient $t$ in the LTE regime. The quantity is given by Eq.~\eqref{eq: evaluation of reflection rate} (or equivalently Eq.~\eqref{eq: evaluation of transfer rate}) and is plotted as a function of $x := m_{\chi}^{\mr{in}}/T_{\chi,f}^{\mr{out}}$ for three benchmark points.
  }
  \label{fig: Plot_rate}
\end{figure}

In the (non-)relativistic limits, the reflection coefficient behaves as
\begin{align}
r \to
\begin{cases}
0, \quad \text{for } v_{\chi,f}^{\mr{out}} \to 1, \\
1, \quad \text{for } v_{\chi,f}^{\mr{out}} \sim \sqrt{T_{\chi, f}^{\mr{out}} / m_{\chi}^{\mr{in}}} \ll 1 ,
\end{cases}
\end{align}
where we have used approximation, ${\pi^4 / 90} \simeq 1$. These asymptotic behaviors can be understood naturally: when the wall velocity is small, the reflectivity $r$ of the DM energy flux is close to unity due to the filtering effect, whereas as the wall velocity increases, the incoming DM can penetrate beyond the wall, so that the reflection efficiency approaches zero. We show in Fig.~\ref{fig: Plot_rate} the velocity dependence of the reflection rate in Eq.~\eqref{eq: evaluation of reflection rate}, where different values of $x := m_{\chi}^{\mr{in}}/T_{\chi,f}^{\mr{out}}$ are chosen as benchmarks.

The parameter $\Delta_r$ can be evaluated as
\begin{align}
\label{eq: estimation of delta r}
\Delta_r 
    \simeq -2 r \left(
        \,\omega_{\chi, r}^{\mr{out}} (\gamma_{\chi, f}^{\mr{out}})^2 (v_{\chi, f}^{\mr{out}})^2 + p_{\chi, f}^{\mr{out}}\right) <0,
\end{align}
if the incident fluid is simply reflected with a reflection coefficient $r$.

\subsubsection{Solution in Ballistic Regime}
The equations to be solved, Eqs.~\eqref{eq: Ballistic matching for energy flux in dark sector} and \eqref{eq: Ballistic matching for momentum in dark sector}, can be rewritten as
\begin{align}
&\tilde{\omega}_{\chi, f}^{\mr{out}} (\gamma_{\chi, f}^{\mr{out}})^2 v_{\chi, f}^{\mr{out}} 
- \omega_{\chi, f}^{\mr{in}} (\gamma_{\chi, f}^{\mr{in}})^2 v_{\chi, f}^{\mr{in}} =0  
        \label{eq: Ballistic matching for energy flux in dark sector 2}\\[0.8em]
&\tilde{\omega}_{\chi, f}^{\mr{out}} (\gamma_{\chi, f}^{\mr{out}})^2 (v_{\chi, f}^{\mr{out}})^2 
	- \omega_{\chi, f}^{\mr{in}} (\gamma_{\chi, f}^{\mr{in}})^2 (v_{\chi, f}^{\mr{in}})^2 \nonumber \\
    &\hspace{2em}
		=   p_{\chi,f }^{\mr{in}} - \tilde{p}_{\chi,f}^{\mr{out}} 
                + \Delta \epsilon_{\chi}^{\mr{eff},r},  
        \label{eq: Ballistic matching for momentum in dark sector 2}
\end{align}
where we have introduced the following parameters:
\begin{align}
\tilde{\omega}_{\chi, f}^{\mr{out}}
    &=\tilde{\rho}_{\chi,f}^{\mr{out}} + \tilde{p}_{\chi,f}^{\mr{out}}, \\
\tilde{\rho}_{\chi,f}^{\mr{out}}
	&:= (1-r)\rho_{\chi,f}^{\mr{out}}, \\
\tilde{p}_{\chi,f}^{\mr{out}}
	&:= (1-r)p_{\chi,f}^{\mr{out}}, \\
\Delta \epsilon_{\chi}^{\mr{eff},r}
	&:= \Delta  \epsilon_{\chi} + \Delta_r .
\end{align}
Here, $\tilde{\rho}_{\chi,f}^{\mr{out}}$ and $\tilde{p}_{\chi,f}^{\mr{out}}$ correspond to the energy density and pressure in which the degrees of freedom are multiplied by a factor of $(1-r)$ outside the wall.

With this parameter replacement, the matching conditions in the presence of reflection reduce to the matching equations without reflection. The solutions to Eqs.~\eqref{eq: Ballistic matching for energy flux in dark sector 2} and \eqref{eq: Ballistic matching for momentum in dark sector 2} can be obtained in almost the same manner as in the template model~\cite{Leitao:2014pda, Giese:2020rtr, Giese:2020znk, Ai:2023see, Sanchez-Garitaonandia:2023zqz}. We review in Appendix~\ref{app: Matching equations without reflection or transfer} the procedure for solving the system without reflection.

For later convenience, we define the pressure difference across the wall as the quantity decomposed as
\begin{align}
\label{eq: difference of p}
\tilde{\Delta} p_{\chi,f}^{r}
	 &:= \tilde{p}_{\chi,f}^{+,r}(T_{\chi,f}^{\mr{out}}) - p_{\chi,f}^{-,r}(T_{\chi, f}^{\mr{in}}) \nonumber \\
	 &= \left(\tilde{p}_{\chi,f}^{+,r}(T_{\chi,f}^{\mr{out}}) 
     - p_{\chi,f}^{-,r}(T_{\chi,f}^{\mr{out}})\right)\nonumber\\
     &\quad
			+ \left(p_{\chi,f}^{-, r}(T_{\chi,f}^{\mr{out}}) - p_{\chi,f}^{-, r}(T_{\chi,f}^{\mr{in}})\right)\\
	&=: D_{r} p_{\chi,f} + \delta p_{\chi,f},
\end{align}
and an analogous definition is adopted for the difference in the energy density.
Here, we have introduced
\begin{align}
&\tilde{\rho}_{\chi,f}^{+,r} = \tilde{\rho}_{\chi,f}^{\mr{out}} + \Delta \epsilon_{\chi}^{\mr{eff},r}, \\
&\tilde{\rho}_{\chi,f}^{-,r} = \tilde{\rho}_{\chi,f}^{\mr{in}}, \\
&\tilde{p}_{\chi,f}^{+, r} = \tilde{p}_{\chi,f}^{\mr{out}} - \Delta \epsilon_{\chi}^{\mr{eff},r}, \\
&\tilde{p}_{\chi,f}^{-, r} = \tilde{p}_{\chi,f}^{\mr{in}}.
\end{align}

The velocity of the DM fluid inside the wall, $v_{\chi, f}^{\mr{in}}$, that satisfies the matching conditions~\eqref{eq: Ballistic matching for energy flux in dark sector 2} and \eqref{eq: Ballistic matching for momentum in dark sector 2} is then given by
\begin{align}
\label{eq: Balestic solution DM}
v_{\chi, f}^{\mr{in}}
	&={1 \over 2 v_{\chi, f}^{\mr{out}}}
        \Bigg[
            G(v_{\chi, f}^{\mr{out}}, \tilde{\alpha}_{\chi,\bar{\theta}}^{r})\nonumber \\
        &\hspace{1em}  
            \pm \sqrt{ G(v_{\chi, f}^{\mr{out}}, \tilde{\alpha}_{\chi,\bar{\theta}}^{r})^2 - 4 (v_{\chi, f}^{\mr{out}})^2 (c_{\chi}^{\mr{in}})^2}
            \Bigg],
\end{align}
where
\begin{align}
\label{eq: G definition}
&G(v, \alpha)
    := (c_{\chi}^{\mr{in}})^2 (1-3 \alpha)
        + v^2 \left(1+ 3 (c_{\chi}^{\mr{in}})^2 \alpha\right),\\
&c_{\chi}^{\mr{in}}
    := \sqrt{\delta p_{\chi,f} \over \delta \rho_{\chi,f}},\\
&\tilde{\alpha}_{\chi,\bar{\theta}}^{r}
    := {1 \over 1-r}\alpha_{\chi,\bar{\theta}}^{r}.
\end{align}
Here, $\alpha_{\chi,\bar{\theta}}^{r}$ is defined in terms of the  pseudotrace $\bar{\theta}_{\chi,f}:= \rho_{\chi,f} - {p_{\chi,f} / (c_{\chi}^{\mr{in}})^2}$, as
\begin{align}
\label{eq: definition of alphachir}
\alpha_{\chi,\bar{\theta}}^{r}
	&:= {D_r \bar{\theta}_{\chi,f} \over 3 \omega_{\chi,f}^{\mr{out}}}, \\
D_r \bar{\theta}_{\chi,f}
    &:= \tilde{\rho}_{\chi,f}^{+,r}(T_{\chi,f}^{\mr{out}}) - \rho_{\chi,f}^{-,r}(T_{\chi,f}^{\mr{out}})\nonumber\\
    &\quad
        -{1 \over (c_{\chi}^{\mr{in}})^2}
            \left(\tilde{p}_{\chi,f}^{+,r}(T_{\chi,f}^{\mr{out}}) 
     - p_{\chi,f}^{-,r}(T_{\chi,f}^{\mr{out}})\right).
\end{align}

Physically, $\alpha_{\chi,\bar{\theta}}^{r}$ corresponds to the dimensionless latent heat if reflection is neglected, i.e., a parameter that characterizes the strength of FOPT. We note that since we have performed a replacement of parameters, $\alpha_{\chi,\bar{\theta}}^{r}$ is, strictly speaking, the dimensionless quantity corresponding to the effective latent heat $\Delta \epsilon_{\chi}^{\mr{eff},r} := \Delta \epsilon_{\chi} + \Delta_r$.
It should also be noted that the parameter appearing in the solution~\eqref{eq: Balestic solution DM} is $\tilde{\alpha}_{\chi,\bar{\theta}}^{r}$, (not $\alpha_{\chi,\bar{\theta}}^{r}$).

For $c_{\chi}^{\mr{in}}$, we adopt the following simplifying approximation:
\begin{align}
\label{eq: sound speed approximation}
{\delta p_{\chi,f} \over \delta \rho_{\chi,f}}
	\simeq {d p_{\chi,f}^{-,r }(T)/dT \over d \rho_{\chi,f}^{-, r}(T)/dT },
\end{align}
and interpret $c_{\chi}^{\mr{in}}$ as the sound speed inside the wall.
In particular, this approximation becomes accurate when the FOPT is not very strong and the temperatures on the two sides of the wall are nearly equal, i.e., $T_{\chi,f}^{\mr{out}} \simeq T_{\chi,f}^{\mr{in}}$~\cite{Giese:2020rtr}.

For the sound speed, we consider two benchmark cases corresponding to different equations of state: a non-relativistic gas with 
$c_{\chi}^{\mr{in}} = \sqrt{T_{\chi,f}^{\mr{out}}/m_{\chi}^{\mr{in}}}$ and a relativistic gas with 
$c_{\chi}^{\mr{in}} = 1/\sqrt{3}$. The former corresponds to a regime in which reflection is efficient, $r \simeq 1$, so that only a small fraction of DM penetrates into the wall; this provides a good approximation when $v_{\chi,f}^{\mr{out}} \ll 1$ (see also Fig.~\ref{fig: Plot_rate}). In contrast, the latter requires that a sufficient amount of DM enters the wall, which is realized only when the reflection is suppressed, $r \simeq 0$, corresponding to the ultra-relativistic limit $v_{\chi,f}^{\mr{out}} \simeq 1$. 

\begin{figure*}[t]
  \centering
  \begin{minipage}[t]{0.48\textwidth}
    \centering
    \includegraphics[width=\textwidth]{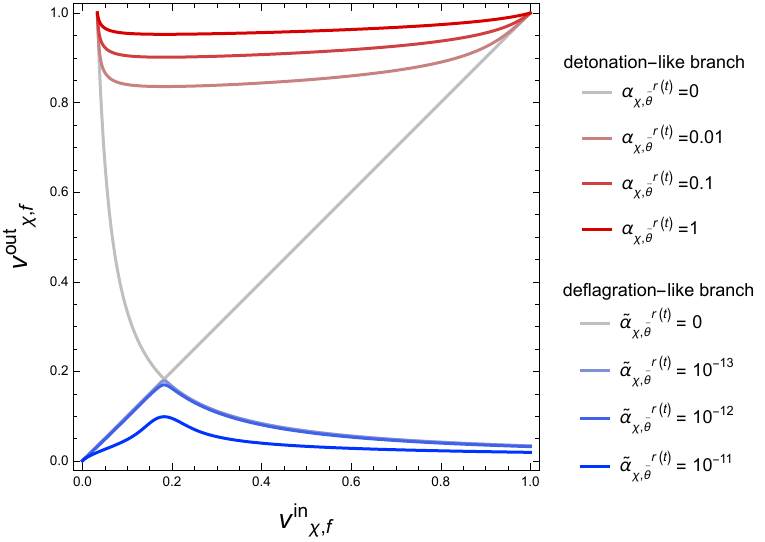}

    \vspace{0.5em}
    \small (left) Non-relativistic gas sound speed $c_{\chi}^{\mr{in}} = \sqrt{T_{\chi,f}^{\mr{out}}/m_{\chi}^{\mr{in}}}$ with the benchmark choice $m_{\chi}^{\mr{in}} / T_{\chi,f}^{\mr{out}} = 30$
  \end{minipage}
  \hfill
  \begin{minipage}[t]{0.48\textwidth}
    \centering
    \includegraphics[width=\textwidth]{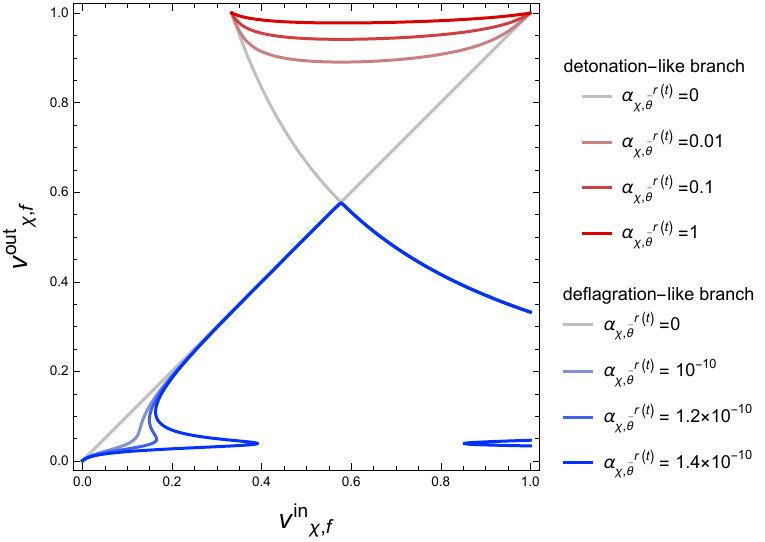}

    \vspace{0.5em}
    \small (right) Relativistic gas sound speed $c_{\chi}^{\mr{in}} = 1/\sqrt{3}$
  \end{minipage}

  \caption{
    The relation between the DM fluid velocities inside and outside the wall obtained from the matching conditions. The relation shown in the figure is valid for both the ballistic and LTE regimes, in the sense that the functional form is the same and the regime dependence enters only through the effective parameter (either $\alpha_{\chi,\bar{\theta}}^{r}$ or $\alpha_{\chi,\bar{\theta}}^{t}$). The curves illustrate the behavior of the solutions for several choices of this parameter.  
    In both panels, the realizable velocity configurations are restricted to the cases where $(v_{\chi,f}^{\mr{out}}, v_{\chi,f}^{\mr{in}})$ are either both supersonic or both subsonic~\cite{Steinhardt:1981ct, Landau:1989, Laine:1993ey, Leitao:2014pda}.
  }
  \label{fig: solution for velocities}
\end{figure*}

In Fig.~\ref{fig: solution for velocities}, we show the solution~\eqref{eq: Balestic solution DM} in the ballistic regime, with the sound speed set to that of a non-relativistic gas (left panel) and a relativistic gas (right panel). To plot these figures, we adopt the benchmark value $m_{\chi}^{\mr{in}} / T_{\chi,f}^{\mr{out}} = 30$ for the ratio of the DM mass to the temperature that determines the strength of the filtering effect. In Appendix~\ref{app: Matching equations without reflection or transfer}, we present figures showing the solutions to the matching conditions in the limit where the reflection effect can be completely neglected.

We note that the realizable velocity configurations shown in the figure are restricted to the cases where $(v_{\chi, f}^{\mr{out}}, v_{\chi, f}^{\mr{in}})$ are either both supersonic or both subsonic~\cite{Steinhardt:1981ct, Landau:1989, Laine:1993ey,Leitao:2014pda}.

Before proceeding, let us first comment on the general features that are common to the two panels.
First, the two branches of the solution, corresponding to positive and negative signs, represent the regimes where $v_{\chi, f}^{\mr{out}}$ is supersonic ($-$) and subsonic ($+$), respectively. Following the standard terminology, we refer to the former as the detonation-like branch and the latter as the deflagration-like branch.
The latter solution exists only when $\tilde{\alpha}_{\chi,\bar{\theta}}^r$ (not $\alpha_{\chi,\bar{\theta}}^r$) is smaller than the squared sound speed outside the wall, $1/3$.\footnote{
A simple way to check this is to verify that $v_{\chi, f}^{\mr{in}} > 0$ in the limit $v_{\chi, f}^{\mr{out}} \ll 1$.
}
In addition, since
\begin{align}
\tilde{\alpha}_{\chi,\bar{\theta}}^{r} 
    = {1 \over 1-r} \alpha_{\chi,\bar{\theta}}^{r}
    \geq \alpha_{\chi,\bar{\theta}}^{r},
\end{align}
the velocity profile typically becomes closer to that of a stronger FOPT compared to the case where reflection is neglected.
This enhancement becomes more significant when $v_{\chi, f}^{\mr{out}}$ is not relativistic, which yields $r \simeq 1$.
The reason is that the reflection effect depends on $v_{\chi, f}^{\mr{out}}$, and the filtering effect becomes more efficient at lower velocities, as shown in Fig~\ref{fig: Plot_rate}.
In particular, the enhancement is pronounced for the deflagration-like branch, where the fluid velocity outside the wall is small.

In addition, we note that
\begin{align}
D_r \bar{\theta}_{\chi,f}
	&=\tilde{\bar{\theta}}_{\chi,f}^{\mr{out}}(T_{\chi,f}^{\mr{out}})- \bar{\theta}_{\chi,f}^{\mr{in}}(T_{\chi,f}^{\mr{out}})\nonumber \\
	&\quad
		+\left(1 + (1/c_{\chi}^{\mr{in}})^2\right)\Delta \epsilon_{\chi}^{\mr{eff},r},
\end{align}
where we have defined $\tilde{\bar{\theta}}_{\chi, f} := (1/(1-r)) \bar{\theta}_{\chi, f}$. The first line on the right-hand side vanishes for $r=0$, while for $r\simeq 1$ each term becomes $\mathcal{O}(1-r)$ and is therefore highly suppressed.
Thus, one typically obtains
\begin{align}
\label{eq: Dr evaluation}
D_r \bar{\theta}_{\chi,f}
	&\simeq \left(1 + (1/c_{\chi}^{\mr{in}})^2\right)\Delta \epsilon_{\chi}^{\mr{eff},r}.
\end{align}

Therefore, independent of the value of the sound speed inside the wall, the existence of the deflagration-like branch is subject to a strong constraint, because we need
\begin{align}
\tilde{\alpha}_{\chi,\bar{\theta}}^{r}
    \simeq {1 \over 1-r}\left(1 + (1/c_{\chi}^{\mr{in}})^2\right)
        {\Delta \epsilon_{\chi}^{\mr{eff},r} \over 3 \omega_{\chi,f}^{\mr{out}}} < {1 \over 3}.
\end{align}
Taking into account the enhancement factor $1/(1-r) \gg 1$, which arises when the wall is not relativistic, this implies that these deflagration-like branches are allowed only when
\begin{align}
\label{eq: deflagration condition ballistic}
\Delta \epsilon_{\chi}^{\mr{eff},r} = \Delta \epsilon_{\chi} + \Delta_r \simeq 0,
\end{align}
namely, when the original latent heat is balanced by the reflected momentum with high precision.

It is also worth noting that, even for the detonation-like branch, the solution is shifted toward larger fluid velocities. This implies that $\tilde{\alpha}_{\chi,\bar{\theta}}^r$ is enhanced relative to $\alpha_{\chi,\bar{\theta}}^r$ by the reflection coefficient. (See also Fig~\ref{fig: solution for velocities wo} in Appendix~\ref{app: Matching equations without reflection or transfer}.)  This is because, for sufficiently large values of $x := m_{\chi}^{\mr{in}}/T_{\chi,f}^{\mr{out}}$, the reflection effect remains significant even for $v_{\chi,f}^{\mr{out}}$ close to unity, as shown in Fig.~\ref{fig: Plot_rate}. Therefore, even in the detonation-like branch, if the condition~\eqref{eq: deflagration condition ballistic} is not satisfied, only solutions with very large fluid velocities for which the reflection effect is negligible can be realized. In this sense, the condition~\eqref{eq: deflagration condition ballistic} can be interpreted as a criterion that determines whether the reflection effect can controls the wall velocity.

Finally, we comment on a characteristic feature of the right panel in Fig.~\ref{fig: solution for velocities}. In this panel, we observe a peculiar behavior: the solution becomes strongly distorted near the origin.
This occurs because the reflection-induced enhancement of $\tilde{\alpha}_{\chi,\bar{\theta}}^r$ depends sensitively on $v_{\chi, f}^{\mr{out}}$, especially when $v_{\chi, f}^{\mr{out}}$ is non-relativistic.
As a consequence, the region near $v_{\chi, f}^{\mr{out}} \simeq 0$ behaves effectively as if the latent heat were much larger, while the solution gradually approaches the small--latent-heat behavior as $v_{\chi, f}^{\mr{out}}$ increases. However, since the choice $c_{\chi}^{\mr{in}} = 1/\sqrt{3}$ is not appropriate for the Filtered DM setup, it is unlikely that this distorted behavior is realized in a physically relevant situation.

\subsubsection{Friction in Ballistic Regime}

In determining the wall velocity, an important condition is the balance between the driving force, given by the difference in the effective potential across the wall, and the backreaction (friction) exerted by the surrounding fluid.
In the ballistic regime, as already studied in the literature~\cite{Gouttenoire:2021kjv, GarciaGarcia:2022yqb, Ai:2024btx}, the reflected component provides an important contribution to the friction.
In the present notation, the backreaction force per unit area is given by
\begin{align}
\label{eq: backreaction force}
\frac{F_{\rm back}}{A}
	&=
        \Delta \left(\omega_{\chi, f} (\gamma_{\chi, f})^2 (v_{\chi, f})^2\right)\nonumber \\
    &\quad
        - r \left(
        \,\omega_{\chi, r}^{\mr{out}} (\gamma_{\chi, f}^{\mr{out}})^2 (v_{\chi, f}^{\mr{out}})^2 + p_{\chi, f}^{\mr{out}}\right) - \Delta_r \nonumber\\
    &\simeq
        \Delta \left(\omega_{\chi, f} (\gamma_{\chi, f})^2 (v_{\chi, f})^2\right) - {1 \over 2} \Delta_r,
\end{align}
where
\begin{align}
&\Delta \left(\omega_{\chi, f} (\gamma_{\chi, f})^2 (v_{\chi, f})^2\right)\nonumber\\
    &\quad 
        := \omega_{\chi, f}^{\mr{out}} (\gamma_{\chi, f}^{\mr{out}})^2 (v_{\chi, f}^{\mr{out}})^2 
	- \omega_{\chi, f}^{\mr{in}} (\gamma_{\chi, f}^{\mr{in}})^2 (v_{\chi, f}^{\mr{in}})^2,
\end{align}
and on the second equality in Eq.~\eqref{eq: backreaction force}, we have used Eq.~\eqref{eq: estimation of delta r}.

To realize a stable solution, this must be balanced with the driving force~\cite{BarrosoMancha:2020fay, Balaji:2020yrx, Ai:2021kak}
\begin{align}
\label{eq: deriving force}
{F_{\mr{pressure}} \over A} 
                    &= \Delta V_{\mr{eff}}\\
					&= p_{\chi,f }^{\mr{in}} -  p_{\chi,f}^{\mr{out}} + \Delta \epsilon_{\chi},
\end{align}
where $\Delta V_{\mr{eff}}$ is the effective potential of the scalar field.

In the limit where the wall velocity is large, one expects not only $r \to 0$ but also $\Delta_r \to 0$, so that the condition reduces to the equilibrium condition in the absence of reflection. On the other hand, when the wall velocity is small, it is necessary to take into account the effect of the additional term, $-(1/2)\Delta_r$. Paying attention to the existence condition of the solution given in Eq.~\eqref{eq: deflagration condition ballistic}, this corresponds to approximately half of the latent heat $\Delta \epsilon$. Therefore, compared to the case where the wall is relativistic, one finds that the corresponding balance condition is equivalent to that with the latent heat reduced by a factor of two.

This balance condition determines the hydrodynamic properties independently of the velocity solution~\eqref{eq: Balestic solution DM}.  Including the stability condition~\eqref{eq: deflagration condition ballistic}, the system is subject to three conditions in total.\footnote{
The balance condition referred to as the ballistic matching condition in the previous work~\cite{Ai:2024btx} corresponds to the stability condition~\eqref{eq: deflagration condition ballistic} in our formulation. The balance between Eqs.~\eqref{eq: backreaction force} and \eqref{eq: deriving force} corresponds to the $z$-component of the energy--momentum matching condition in the previous work~\cite{Ai:2024btx}.
} On the other hand, the number of parameters that characterize the hydrodynamics of the DM fluid is five in the ballistic regime:
\begin{align}
T_{\chi,f}^{\mr{out}(\mr{in})}, \, v_{\chi,f}^{\mr{out}(\mr{in})},\, \Delta \epsilon_{\chi} \, .
\end{align}  
Therefore, once the nucleation temperature during the FOPT and the latent heat parameter $\Delta \epsilon$, both of which can be computed in a specific model, are determined, the hydrodynamics of the system can be solved completely, including the determination of the wall velocity. However, a detailed model-dependent analysis is beyond the scope of this paper.

\subsection{LTE Regime}
Next, we consider the LTE regime in which reflection can be neglected. In this case, even near the wall, the DM fluid and the radiation fluid do not decouple, and it is necessary to take into account the transport of energy and momentum from the DM fluid to the radiation fluid. By integrating both sides of Eqs.~\eqref{eq: energy momentum evolution for dark matter plasma} and \eqref{eq: energy momentum evolution for radiation plasma} with respect to $z$, two sets of matching conditions, each consisting of two equations, are obtained as follows:
\begin{align}
&\omega_{\chi, f}^{\mr{out}} (\gamma_{\chi, f}^{\mr{out}})^2 v_{\chi, f}^{\mr{out}} 
- \omega_{\chi, f}^{\mr{in}} (\gamma_{\chi, f}^{\mr{in}})^2 v_{\chi, f}^{\mr{in}} \nonumber \\
	&\hspace{9em}
        =t \left(\omega_{\chi, f}^{\mr{out}} (\gamma_{\chi, f}^{\mr{out}})^2 v_{\chi, f}^{\mr{out}}\right),  
        \label{eq: LTE matching for energy flux in dark sector}\\[0.8em]
&\omega_{\chi, f}^{\mr{out}} (\gamma_{\chi, f}^{\mr{out}})^2 (v_{\chi, f}^{\mr{out}})^2 
	- \omega_{\chi, f}^{\mr{in}} (\gamma_{\chi, f}^{\mr{in}})^2 (v_{\chi, f}^{\mr{in}})^2 \nonumber \\
    &\hspace{2em}
		=   p_{\chi,f }^{\mr{in}} - (1 - t) p_{\chi,f}^{\mr{out}} 
                + \Delta \epsilon_{\chi, f} +  \Delta_{t}
        \nonumber \\
        &\hspace{5em}
			+ t \left(\omega_{\chi, f}^{\mr{out}} (\gamma_{\chi, f}^{\mr{out}})^2 (v_{\chi, f}^{\mr{out}})^2\right),
        \label{eq: LTE matching for momentum in dark sector}\\[0.8em]
&\omega_{\mr{rad}}^{\mr{out}} (\gamma_{\mr{rad}}^{\mr{out}})^2 v_{\mr{rad}}^{\mr{out}} 
	- \omega_{\mr{rad}}^{\mr{in}} (\gamma_{\mr{rad}}^{\mr{in}})^2 v_{\mr{rad}}^{\mr{in}} \nonumber\\
    &\hspace{9em}
	=- t \left(\omega_{\chi, f}^{\mr{out}} (\gamma_{\chi, f}^{\mr{out}})^2 v_{\chi, f}^{\mr{out}}\right),  \label{eq: LTE matching for energy flux in SM sector}\\[0.8em]
&\omega_{\mr{rad}}^{\mr{out}} (\gamma_{\mr{rad}}^{\mr{out}})^2 (v_{\mr{rad}}^{\mr{out}})^2
	- \omega_{\mr{rad}}^{\mr{in}} (\gamma_{\mr{rad}}^{\mr{in}})^2 (v_{\mr{rad}}^{\mr{in}})^2 \nonumber \\
		 &\hspace{2em}
        =   p_{\mr{rad}}^{\mr{in}} - p_{\mr{rad}}^{\mr{out}} 
            - t p_{\chi}^{\mr{out}}
                + \Delta \epsilon_{\mr{rad}}- \Delta_{t}
        \nonumber \\
        &\hspace{5em}
			- t \left(\omega_{\chi, f}^{\mr{out}} (\gamma_{\chi, f}^{\mr{out}})^2 (v_{\chi, f}^{\mr{out}})^2\right).
    \label{eq: LTE matching for momentum in SM sector}
\end{align}
where
\begin{align}
\omega_{\mr{rad}} := \rho_{\mr{rad}} + p_{\mr{rad}},  
\end{align}
denotes the enthalpy density of the radiation fluid,  while $\rho_{\mr{rad}}$ and $p_{\mr{rad}}$ are the corresponding energy density and pressure, respectively. Although it has been neglected so far, in our present setup, there may exist particles within the radiation fluid that acquire a tiny mass during FOPT. The latent heat associated with such particles is denoted by $\Delta \epsilon_{\mr{rad}}$ in Eq.~\eqref{eq: LTE matching for momentum in SM sector}. Naturally, the difference in the latent heat experienced by the radiation sector is negligibly small compared to the corresponding difference in the latent heat of the DM sector:
\begin{align}
\Delta \epsilon_{\mr{rad}} \ll \Delta \epsilon_{\chi,f}.
\end{align}

Since the radiation component follows an equilibrium distribution both inside and outside the wall, the energy density and pressure can thus be expressed as follows:
\begin{align}
    & \rho_{\mr{rad}}^{\mr{out}(\mr{in})}
	= {\pi^2 \over 30} g_{\mr{rad}*} \left(T_{\mr{rad}}^{\mr{out}(\mr{in})}\right)^4, \\
&  p_{\mr{rad}}^{\mr{out}(\mr{in})}
	= {\pi^2 \over 90} g_{\mr{rad}*} \left(T_{\mr{rad}}^{\mr{out}(\mr{in})}\right)^4 ,
\end{align}
where $g_{\mr{rad}*}$ denotes the effective degrees of freedom for the radiation sector.

The introduced parameters $t$ and $\Delta_t$ represent the fractions of the incident DM fluid energy flux and momentum that are converted into the radiation component through relaxation processes in the vicinity of the wall:
\begin{align}
t
    &:= {- \int dz \,\Theta_{\chi}^0 \over \omega_{\chi, r}^{\mr{out}} (\gamma_{\chi, f}^{\mr{out}})^2 v_{\chi, f}^{\mr{out}} } >0,  \\
\label{eq: anzatz for reflection}
\Delta_t
    &:=
    - \int dz \,\Theta_{\chi}^z 
    - t \left(
        \,\omega_{\chi, r}^{\mr{out}} (\gamma_{\chi, f}^{\mr{out}})^2 (v_{\chi, f}^{\mr{out}})^2 + p_{\chi, f}^{\mr{out}}
        \right).
\end{align}

The parameter $t$ is then given by a calculation analogous to that in the previous section~\eqref{eq: evaluation of reflection rate} as
\begin{align}
\label{eq: evaluation of transfer rate}
t
    &\simeq 1- {F_w \over \omega_{\chi, f}^{\mr{out}} (\gamma_{\chi, f}^{\mr{out}})^2 v_{\chi, f}^{\mr{out}}}, 
\end{align}
and, in the (non-)relativistic limits, it behaves as
\begin{align}
t \to
\begin{cases}
0, \quad \text{for } v_{\chi,f}^{\mr{out}} \to 1, \\
1, \quad \text{for } v_{\chi,f}^{\mr{out}} \sim \sqrt{T_{\chi, f}^{\mr{out}} / m_{\chi}^{\mr{in}}} \ll 1 .
\end{cases}
\end{align}
For the detailed velocity dependence, see Fig.~\ref{fig: Plot_rate}.

If the momentum of the incident DM fluid is transferred to the SM fluid with the same transfer fraction $t$, the parameter $\Delta_t$ can be approximated as
\begin{align}
\label{eq: estimation of delta t}
\Delta_t \simeq 0 \, .
\end{align}
This should be contrasted with the ballistic regime, where $\Delta_r$ gives a sizable contribution (see Eq.~\ref{eq: estimation of delta r}).

\subsubsection{Solution in LTE Regime}

The set of equations to be solved, Eqs.~\eqref{eq: LTE matching for energy flux in dark sector}-\eqref{eq: LTE matching for momentum in SM sector}, can be rewritten as follows:
\begin{align}
&\bar{\omega}_{\chi, f}^{\mr{out}} (\gamma_{\chi, f}^{\mr{out}})^2 v_{\chi, f}^{\mr{out}} 
- \omega_{\chi, f}^{\mr{in}} (\gamma_{\chi, f}^{\mr{in}})^2 v_{\chi, f}^{\mr{in}} 
        =0,  
        \label{eq: LTE matching for energy flux in dark sector 2}\\[0.8em]
&\bar{\omega}_{\chi, f}^{\mr{out}} (\gamma_{\chi, f}^{\mr{out}})^2 (v_{\chi, f}^{\mr{out}})^2 
	- \omega_{\chi, f}^{\mr{in}} (\gamma_{\chi, f}^{\mr{in}})^2 (v_{\chi, f}^{\mr{in}})^2 \nonumber \\
    &\hspace{2em}
		=   p_{\chi,f }^{\mr{in}} - \bar{p}_{\chi,f}^{\mr{out}} 
                + \Delta \epsilon_{\chi, f}^{\mr{eff},t},
        \label{eq: LTE matching for momentum in dark sector 2}\\[0.8em]
&\bar{\omega}_{\mr{rad}}^{\mr{out}} (\gamma_{\mr{rad}}^{\mr{out}})^2 v_{\mr{rad}}^{\mr{out}} 
	- \omega_{\mr{rad}}^{\mr{in}} (\gamma_{\mr{rad}}^{\mr{in}})^2 v_{\mr{rad}}^{\mr{in}} 
	=0,  \label{eq: LTE matching for energy flux in SM sector 2}\\[0.8em]
&\bar{\omega}_{\mr{rad}}^{\mr{out}} (\gamma_{\mr{rad}}^{\mr{out}})^2 (v_{\mr{rad}}^{\mr{out}})^2
	- \omega_{\mr{rad}}^{\mr{in}} (\gamma_{\mr{rad}}^{\mr{in}})^2 (v_{\mr{rad}}^{\mr{in}})^2 \nonumber \\
		 &\hspace{2em}
        =   p_{\mr{rad}}^{\mr{in}} - \bar{p}_{\mr{rad}}^{\mr{out}} 
                + \Delta \epsilon_{\mr{rad}}^{\mr{eff},t}.
    \label{eq: LTE matching for momentum in SM sector 2}
\end{align}
where we have introduced the following parameters:
\begin{align}
\bar{\rho}_{\chi,f}^{\mr{out}}
	&:= (1-t)\rho_{\chi,f}^{\mr{out}}, \\
\bar{p}_{\chi,f}^{\mr{out}}
	&:= (1-t)p_{\chi,f}^{\mr{out}}, \\
\bar{\rho}_{\mr{rad}}^{\mr{out}}
    &:= \left(1 + t {g_{\chi} \over g_{\mr{rad}*}}\right) \rho_{\mr{rad}}^{\mr{out}},\\
\bar{p}_{\mr{rad}}^{\mr{out}}
    &:= \left(1 + t {g_{\chi} \over g_{\mr{rad*}}}\right) p_{\mr{rad}}^{\mr{out}}, \\
\Delta \epsilon_{\chi}^{\mr{eff},t}
	&:= \Delta  \epsilon_{\chi} + \Delta_t,\\
\Delta \epsilon_{\mr{rad}}^{\mr{eff},t}
	&:= \Delta  \epsilon_{\mr{rad}} - \Delta_t.
\end{align}
Here, $\bar{\rho}_{\chi,f}^{\mr{out}}$ and $\bar{p}_{\chi,f}^{\mr{out}}$ denote the DM energy density and pressure with the effective degrees of freedom multiplied by a factor of $(1-t)$, while $\bar{\rho}_{\mr{rad}}^{\mr{out}}$ and $\bar{p}_{\mr{rad}}^{\mr{out}}$ represent the radiation energy density and pressure with the effective degrees of freedom rescaled by a factor of $(1 + t\, g_{\chi}/g_{\mr{rad}})$. Since typically $g_{\chi} \ll g_{\mr{rad}*}$, the effect experienced by the radiation fluid across the wall is much smaller than that for the DM fluid.

Similarly to the ballistic regime, in the LTE regime, it is also convenient to introduce the following decomposition
\begin{align}
\label{eq: difference of p}
\bar{\Delta} p_{\chi,f}^{t}
	 &:= \bar{p}_{\chi,f}^{+,t}(T_{\chi,f}^{\mr{out}}) - p_{\chi,f}^{-,t}(T_{\chi, f}^{\mr{in}}) \nonumber \\
	 &= \left(\bar{p}_{\chi,f}^{+,t}(T_{\chi,f}^{\mr{out}}) 
     - p_{\chi,f}^{-,t}(T_{\chi,f}^{\mr{out}})\right)\nonumber\\
     &\quad
			+ \left(p_{\chi,f}^{-, t}(T_{\chi,f}^{\mr{out}}) - p_{\chi,f}^{-, t}(T_{\chi,f}^{\mr{in}})\right)\\
	&=: D_{t} p_{\chi,f} + \delta p_{\chi,f},
\end{align}
and the same decomposition applies to the difference in the energy density. Here, we also introduce the following parameters:
\begin{align}
&\bar{\rho}_{\chi,f}^{+,t} =  
    \bar{\rho}_{\chi,f}^{\mr{out}} + \Delta \epsilon_{\chi}^{\mr{eff},t}, \\
&\bar{\rho}_{\chi,f}^{-,t} 
    = \bar{\rho}_{\chi,f}^{\mr{in}}, \\
&\bar{p}_{\chi,f}^{+, t} = 
    \bar{p}_{\chi,f}^{\mr{out}} - \Delta \epsilon_{\chi}^{\mr{eff},t}, \\
&\bar{p}_{\chi,f}^{-, t} 
    = \bar{p}_{\chi,f}^{\mr{in}},
\end{align}
and also an analogous definition is adopted for the radiation sector.

With this factorization, the two equations for the DM fluid and the two equations for the radiation fluid become decoupled. Moreover, each set reduces to the matching conditions in the absence of transfer effects. Therefore, as in the previous section, the solutions to Eqs.~\eqref{eq: LTE matching for energy flux in dark sector 2}--\eqref{eq: LTE matching for momentum in SM sector 2} can be obtained in essentially the same manner as in the template model~\cite{Leitao:2014pda, Giese:2020rtr, Giese:2020znk, Ai:2023see, Sanchez-Garitaonandia:2023zqz}.

The velocity of the DM fluid inside the wall, $v_{\chi,f}^{\mr{in}}$, which satisfies the matching conditions~\eqref{eq: LTE matching for energy flux in dark sector 2} and \eqref{eq: LTE matching for momentum in dark sector 2}, and the velocity of the radiation fluid inside the wall, $v_{\mr{rad}}^{\mr{in}}$, which satisfies the matching conditions~\eqref{eq: LTE matching for energy flux in SM sector 2} and \eqref{eq: LTE matching for momentum in SM sector 2}, are respectively given by
\begin{align}
\label{eq: LET solution DM}
v_{\chi, f}^{\mr{in}}
	&={1 \over 2 v_{\chi, f}^{\mr{out}}}
        \Bigg[
            G(v_{\chi, f}^{\mr{out}}, \tilde{\alpha}_{\chi,\bar{\theta}}^{t})\nonumber \\
        &\hspace{1em}  
            \pm \sqrt{ G(v_{\chi, f}^{\mr{out}}, \tilde{\alpha}_{\chi,\bar{\theta}}^{t})^2 - 4 (v_{\chi, f}^{\mr{out}})^2 (c_{\chi}^{\mr{in}})^2}
            \Bigg], \\
\label{eq: LET solution radiation}
v_{\mr{rad}}^{\mr{in}}
	&={1 \over 2 v_{\mr{rad}}^{\mr{out}}}
        \Bigg[
            G(v_{\mr{rad}}^{\mr{out}}, \tilde{\alpha}_{\mr{rad},\bar{\theta}}^{t})\nonumber \\
        &\hspace{1em}  
            \pm \sqrt{ G(v_{\mr{rad}}^{\mr{out}}, \tilde{\alpha}_{\mr{rad},\bar{\theta}}^{t})^2 - (4/3) (v_{\mr{rad}}^{\mr{out}})^2 }
            \Bigg],
\end{align}
where
\begin{align}
&\tilde{\alpha}_{\chi,\bar{\theta}}^{t}
    := {1 \over 1-t}\alpha_{\chi,\bar{\theta}}^{t}, \\
&\tilde{\alpha}_{\mr{rad}}^{t}
    := {1 \over 1+t(g_{\chi}/g_{\mr{rad}*})}\alpha_{\mr{rad},\bar{\theta}}^{t}.
\end{align}
As in the previous section, $\alpha_{\chi,\bar{\theta}}^{t}$ is defined in terms of the pseudotrace $\bar{\theta}_{\chi, f}$ while $\tilde{\alpha}_{\mr{rad}}^{t}$ is defined by the trace $\theta_{\mr{rad}}:= \rho_{\mr{rad}} - 3 p_{\mr{rad}}$ as
\begin{align}
\alpha_{\chi,\bar{\theta}}^{t}
	:= {D_t \bar{\theta}_{\chi,f} \over 3 \omega_{\chi,f}^{\mr{out}}},\\
\alpha_{\mr{rad}}^{t}
	:= {D_t \theta_{\mr{rad}} \over 3 \omega_{\mr{rad}}^{\mr{out}}},
\end{align}
with
\begin{align}
D_t \bar{\theta}_{\chi,f}
    &:= \bar{\rho}_{\chi,f}^{+,t}(T_{\chi,f}^{\mr{out}}) - \rho_{\chi,f}^{-,t}(T_{\chi,f}^{\mr{out}})\nonumber\\
    &\quad
        -{1 \over (c_{\chi}^{\mr{in}})^2}
            \left(\bar{p}_{\chi,f}^{+,t}(T_{\chi,f}^{\mr{out}}) 
     - p_{\chi,f}^{-,t}(T_{\chi,f}^{\mr{out}})\right),\\
D_t \theta_{\mr{rad}}
    &:= \bar{\rho}_{\mr{rad}}^{+,t}(T_{\mr{rad}}^{\mr{out}}) - \rho_{\mr{rad}}^{-,t}(T_{\mr{rad}}^{\mr{out}})\nonumber\\
    &\quad
        -3
            \left(\bar{p}_{\mr{rad}}^{+,t}(T_{\mr{rad}}^{\mr{out}}) 
     - p_{\mr{rad}}^{-,t}(T_{\mr{rad}}^{\mr{out}})\right).
\end{align}
As for the sound speed inside the wall, similarly to the ballistic regime, we adopted the expression obtained by replacing $r$ with $t$ in the relation~\eqref{eq: sound speed approximation}. This provides a good approximation when the temperature variation across the wall is small.

For the same reason as discussed in the previous section, $D_t \bar{\theta}_{\chi,f}$ can be estimated as
\begin{align}
D_t \bar{\theta}_{\chi,f}
	&\simeq \left(1 + (1/c_{\chi}^{\mr{in}})^2\right)\Delta \epsilon_{\chi}^{\mr{eff},t}, \\
D_t \bar{\theta}_{\mr{rad}}
	&\simeq 4 \Delta \epsilon_{\mr{rad}}^{\mr{eff},t}.
\end{align}

Compared with the ballistic regime, the difference in the expression of the solution for the DM fluid is encapsulated in the effective parameter (either $\tilde{\alpha}_{\chi,\bar{\theta}}^{r}$ or $\tilde{\alpha}_{\chi,\bar{\theta}}^{t}$). Therefore, the behavior of Eq.~\eqref{eq: LET solution DM} is also described by Figs.~\ref{fig: solution for velocities}, depending on the sound speed.

By analogy with the discussion in the ballistic regime, the deflagration-like branch is allowed as a solution only if
\begin{align}
\label{eq: deflagration condition LTE}
\Delta \epsilon_{\chi}^{\mr{eff},t} = \Delta \epsilon_{\chi} + \Delta_t \simeq 0,
\end{align}
is satisfied. However, unlike in the ballistic regime, in the LTE regime one expects $\Delta_t \simeq 0$ (see Eq.~\eqref{eq: estimation of delta t}). Therefore, Eq.~\eqref{eq: deflagration condition LTE} requires a very small latent heat. Consequently, compared with the ballistic regime, the parameter region in which the deflagration-like branch can be realized is more restricted in the LTE regime.

On the other hand, for the radiation fluid, the effect of the transfer is negligibly small due to the suppression factor $g_{\chi}/g_{\mr{rad}*}$, and $\alpha_{\mr{rad}}^{t}$ itself is small. Therefore, the impact on the fluid induced by crossing the wall is limited.

\subsubsection{Friction in LTE Regime}
Following the ballistic regime, we examine the balance between the back reaction (friction) from the fluid and the driving force, which is crucial for determining the terminal wall velocity.

In the LTE regime, reflection can be neglected, and thus the friction is determined by the change in momentum as the fluid crosses the wall~\cite{BarrosoMancha:2020fay, Balaji:2020yrx, Ai:2021kak}. The total friction is given by the sum of the contributions from the DM fluid and the radiation fluid:\footnote{
We note that the change in the radiation momentum is small in the following sense:
\begin{align}
{\Delta \left(\omega_{\mr{rad}} (\gamma_{\mr{rad}})^2 v_{\mr{rad}}\right) \over \omega_{\mr{rad}}^{\mr{out}} (\gamma_{\mr{rad}}^{\mr{out}})^2 v_{\mr{rad}}^{\mr{out}} } \sim t {g_{\chi} \over g_{\mr{rad}*}} \ll 1.
\end{align}
However, this does not imply that it can be neglected when compared with $\omega_{\chi, f}^{\mr{out}} (\gamma_{\chi, f}^{\mr{out}})^2 v_{\chi, f}^{\mr{out}}$.
}
\begin{align}
\frac{F_{\rm back}}{A}
	&= \Delta \left(\omega_{\chi, f} (\gamma_{\chi, f})^2 (v_{\chi, f})^2\right)
	 	+ \Delta \left(\omega_{\mr{rad}} (\gamma_{\mr{rad}})^2 (v_{\mr{rad}})^2\right).
\end{align}
Since we have assumed that no energy and momentum is exchanged with the wall during the transfer process, the origin of the friction can be classified into two contributions: the momentum change due to particle-level deceleration associated with the DM fluid acquiring mass~\cite{Bodeker:2009qy, Bodeker:2017cim}, and the momentum change corresponding to the variation of temperature (and velocity) across the wall at the fluid level~\cite{BarrosoMancha:2020fay, Balaji:2020yrx, Ai:2021kak}.  
When the transfer rate is close to unity, almost no DM fluid can penetrate into the inside of the wall. Therefore, unless the fluid velocity is highly relativistic, the former contribution can be neglected, and the friction is largely determined by the latter.

The driving force to be balanced is given by
\begin{align}
\frac{F_{\mr{pressure}}}{A} 
        &= 
            \Delta V_{\mr{eff}}\\
		&= 
            p_{\chi,f}^{\mr{in}} - p_{\chi,f}^{\mr{out}} + \Delta \epsilon_{\chi}, \nonumber\\
        &\qquad
            + p_{\mr{rad}}^{\mr{in}} - p_{\mr{rad}}^{\mr{out}} 
                + \Delta \epsilon_{\mr{rad}}.           
\end{align}
For the deflagration branch to exist, a very small $\Delta \epsilon_{\chi}$ is required. Under this condition, a stable solution can exist only when the friction is sufficiently small, which in turn requires that both the temperature change and the velocity change of the fluid across the wall are very small, or when the difference in the friction from the DM sector is canceled by radiation carrying a pressure difference with the opposite sign. 

In the LTE regime, the number of parameters underlying the hydrodynamic description increases to ten:
\begin{align}
T_{\chi,f(\mr{rad})}^{\mr{out}(\mr{in})}, \, v_{\chi,f(\mr{rad})}^{\mr{out}(\mr{in})},\, \Delta \epsilon_{\chi(\mr{rad})} \, .
\end{align}
On the other hand, there are four matching conditions~\eqref{eq: LTE matching for energy flux in dark sector}--\eqref{eq: LTE matching for momentum in SM sector}. Including the existence condition for the solution~\eqref{eq: deflagration condition LTE} and the boundary conditions outside the wall,~\eqref{eq: assumption outside wall temperature} and \eqref{eq: assumption outside wall velocity}, one obtains seven constraint equations in total. Therefore, once the nucleation temperature and the latent heats $\Delta \epsilon_{\chi(\mr{rad})}$ are computed in a specific model and provided as inputs, the system of equations can be solved exactly. However, a detailed model-dependent analysis is beyond the scope of this paper.

\subsubsection{Negative entropy production}
Finally, we comment on an interesting property related to entropy production that emerges from the analysis of two-component hydrodynamics in the LTE regime. As stated in Eq.~\eqref{eq: entropy non-conservation in LTE} in the previous section, even in the LTE regime, once energy transfer is taken into account, the entropy current is not conserved for each component, and moreover, it is not conserved even for the total fluid.

Again, adopting the approximation that the $z$-direction is aligned with the wall expansion, the nontrivial components of the entropy current evolution equations are those related to
\begin{align}
\partial_{z} S_{\mr{rad}(\chi)}^{z}(z).
\end{align}
We define the integrated quantities with respect to $z$ as
\begin{align}
\delta \sigma_{\chi,f }&:=\int dz\,\partial_{z}S_{\chi,f}^{z}, \\
\delta \sigma_{\mr{rad}}&:=\int dz\,\partial_{z}S_{\mr{rad}}^{z},
\label{eq: global entropy production in integral}
\end{align}
and define their sum as $\delta \sigma := \delta \sigma_{\chi,f }+\delta \sigma_{\mr{rad}}$.

For simplicity, we consider the case where the wall is sufficiently slow. In this limit, the enthalpy of the DM fluid inside the wall, $\omega_{\chi,f}^{\mr{in}}$, as well as the number density $n_{\chi,f}^{\mr{in}}$, are negligibly small. Under this simplification, evaluating $\delta \sigma$ from its definition to linear order in the fluid velocities, we obtain
\begin{align}
\label{eq: global entropy production}
\delta \sigma
   	 &\simeq  
	- \frac{\omega_{\chi,f}^{\mr{out}}}{T_{\chi,f}^{\mr{out}}} v_{\chi,f}^{\mr{out}}
    	+ \frac{\omega_{\mr{rad}}^{\mr{in}}}{T_{\mr{rad}}^{\mr{in}}} v_{\mr{rad}}^{\mr{in}} 
 	-\frac{\omega_{\mr{rad}}^{\mr{out}}}{T_{\mr{rad}}^{\mr{out}}} v_{\mr{rad}}^{\mr{out}}, \\
	&\simeq  \omega_{\mr{rad}}^{\mr{in}} v_{\mr{rad}}^{\mr{in}} \left({1 \over T_{\mr{rad}}^{\mr{in}}}  
			- {1 \over T_{\mr{rad}}^{\mr{out}}} \right).
\end{align}
where we assumed that the DM mass can be neglected outside the wall, $\mu_{\chi}^{\mr{out}} \simeq 0$. In the second equality, we used the relation that the temperatures and velocity of the two fluids are equal outside the wall (see Eqs.~\eqref{eq: assumption outside wall temperature} and \eqref{eq: assumption outside wall velocity}), together with energy-flux conservation obtained by summing the matching conditions for the two components, which implies
\begin{equation}
\omega_{\mr{\chi},f}^{\mr{out}} v_{\chi,f}^{\mr{out}}+ \omega_{\mr{rad}}^{\mr{out}}v_{\mr{rad}}^{\mr{out}} \simeq \omega_{\mr{rad}}^{\mr{in}} v_{\mr{rad}}^{\mr{in}}.
\end{equation}

From Eq.~\eqref{eq: global entropy production}, if the radiation temperature increases inside the wall—this is rather a natural situation when the energy-momentum of dark matter is injected into radiation—negative entropy production occurs,
\begin{align}
\delta \sigma < 0.
\end{align}
One might think that this consequence violates the second law of thermodynamics. This is because the entropy current $S^{\mu}$ is interpreted such that its corresponding charge is the entropy density, while its spatial components represent the entropy flux (or a dimensionless heat flux). From this viewpoint, a negative evolution of this quantity appears peculiar from the standpoint of the second law, for which one usually requires that the derivative of the entropy current should be non-negative~\cite{Balaji:2020yrx, Ai:2021kak}.

However, we note that this negative entropy production does not violate the second law of thermodynamics for the total system, because the free energy of the full system, including the scalar field, always decreases; otherwise, the FOPT cannot proceed. 

Then, it may appear counterintuitive that the second law seems to be violated at the level of a subsystem, even though it is satisfied for the total system.  This situation may be interpreted from the viewpoint of information thermodynamics~\cite{Brillouin:1951, Bennett:1982, Landauer:1961, Dine:2003ax, Morrissey:2012db}.

In information-thermodynamic systems, as exemplified by Maxwell's demon~\cite{Maxwell:1871}, negative entropy production can occur when measurements and the feedback based on the measurement act on the system~\cite{Brillouin:1951, Bennett:1982, Landauer:1961, Dine:2003ax, Morrissey:2012db}. However, once the free energy of the demon, which records the measurement outcomes and implements the feedback, is taken into account, the second law is restored for the total system~\cite{Shizume:1995, Piechocinska:2000,  Touchette:2000, Touchette:2004,  Barkeshli:2005, Kim:2007, Sagawa:2008bsv, Dillenschneider:2009, Turgut:2009, Sagawa:2009jba, Horowitz:2010, Reeb:2014drq, Sagawa:2012, Sagawa:2012pre, Abreu:2012, Lahiri:2012, Still:2012ta, Sagawa:2012prl, Sagawa:2013, Sagawa:2014}.

In the present system, the wall microscopically ``measures'' the particle species and provides ``feedback'' depending on the outcome: radiation particles are allowed to pass, while DM particles are strongly restricted from entering and are forced to relax with the radiation sector. Therefore, at least when viewed in the rest frame of the wall, this behavior closely resembles that considered in information thermodynamics. In fact, the Filtered DM scenario has properties very similar to those of a semipermeable membrane, and semipermeable membranes exhibit behavior closely analogous to that of Maxwell's demon~\cite{Maruyama:2009zza}.

A more detailed investigation along this line of interpretation is left for future work.
\section{Impact of DM Relic Density}
\label{sec: Impact of DM Relic Density}
Building on the discussion so far, we revisit how hydrodynamic effects influence the relic density of Filtered DM, as discussed in earlier work~\cite{Jiang:2023nkj}.

In each of the ballistic and LTE regimes, we consider the number density that can penetrate into the interior of the wall, which becomes a dark matter relic in the present Universe. For solutions on the deflagration branch, the wall velocity is identified with the fluid velocity inside the wall, $v_w = v_{\chi,f}^{\mr{in}}$. The same applies to the nucleation temperature, $T_n = T_{\chi,f}^{\mr{in}}$.\footnote{For the detonation-like branch, by contrast, we identify the wall velocity and the nucleation temperature with those of the fluid outside the wall: $v_w = v_{\chi,f}^{\mr{out}}$ and $T_n = T_{\chi,f}^{\mr{out}}$. In this case as well, the number density that can penetrate into the interior of the wall is affected by the change in velocity across the wall. However, for the detonation-like branch, in which the wall velocity is high, the filtering effect is expected not to operate efficiently. We therefore focus on the deflagration branch in the following analysis.}

On the other hand, since the filtering effect is determined by the velocity of the fluid incident from outside the wall, the number flux penetrating the wall can be expressed in terms of the velocity and temperature outside it. The number density of particles that can enter the wall is therefore written as follows:
\begin{align}
\label{eq: number density inside hydro}
n_{\chi}^{\mr{hyd},\mr{in}}  
        &\simeq
	       \frac{g_{\chi} (T_{\chi,f}^{\mr{out}})^3}{\gamma_{\chi,f}^{\mr{in}} v_{\chi,f}^{\mr{in}}}
	       \left(
	        \frac{\gamma_{\chi, f}^{\mr{out}}(1-v_{\chi,f}^{\mr{out}})m_{\chi}^{\mr{in}}/T_{\chi,f}^{\mr{out}} + 1}{4 \pi^2 (\gamma_{\chi,f}^{\mr{out}})^3 (1-v_{\chi,f}^{\mr{out}})^2}
	        \right) \nonumber \\
    &\qquad
        \times
	       e^{- \gamma_{\chi,f}^{\mr{out}} (1-v_{\chi,f}^{\mr{out}}) m_{\chi}^{\mr{in}}/T_{\chi,f}^{\mr{out}} } .
\end{align}
Hereafter, we assume that the change in the fluid temperature across the wall is sufficiently small, so that the approximation in Eq.~\eqref{eq: sound speed approximation} becomes accurate.
Comparing Eq.~\eqref{eq: number density inside hydro} with Eq.~\eqref{eq: number density inside}, one finds that the number density that can enter the wall is modified by the change in velocity across the wall, relative to the case in which this effect is neglected.

We note that in the ballistic case, there additionally remains a reflected component outside the wall. This component may eventually collapse to form primordial black holes~\cite{Baker:2021nyl, Jung:2021mku, Gehrman:2023qjn}, and may affect the final DM abundance. However, in the following, we do not include this contribution in the DM relic abundance.

\begin{figure}[t]
  \centering
  {\includegraphics[width=0.47\textwidth]{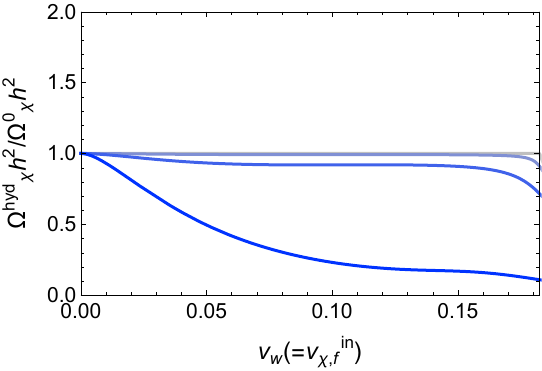}} 
  \caption{
  The impact of hydrodynamics on the DM abundance is expressed as the ratio to the DM abundance obtained when hydrodynamics is neglected. The benchmark values of the effective latent heat parameter $\alpha_{\chi,\bar{\theta}}^{r(t)}$ are taken to be the same as in the left panel in Fig.~\ref{fig: solution for velocities}: from the lightest to the darkest color, $\alpha_{\chi,\bar{\theta}}^{r(t)} = 0, 10^{-13}, 10^{-12}, 10^{-11}$. The sound speed is taken to be that of a non-relativistic gas, $c_{\chi}^{\mr{in}} = \sqrt{T_{\chi,f}^{\mr{out}}/m_{\chi}^{\mr{in}}}$, with  $m_{\chi}^{\mr{in}} / T_{\chi,f}^{\mr{out}} = 30$. The temperature variation is assumed to be small in the DM sector, and is negligibly small in the radiation sector.
  }
  \label{fig: DM Abundance}
\end{figure}

We denote the DM abundance obtained from $n_{\chi}^{\mr{hyd},\mr{in}}$ by $\Omega_{\chi}^{\mr{hyd}} h^2$. To quantify hydrodynamic effects, Fig.~\ref{fig: DM Abundance} shows the ratio to the DM abundance without hydrodynamics, Eq.~\eqref{eq: Filtered DM abundance},
\begin{align}
\frac{\Omega_{\chi}^{\mr{hyd}} h^2}{\Omega_{\chi} h^2}.
\end{align}

In Fig.~\ref{fig: DM Abundance},  we display only the deflagration branch since a non-relativistic wall is preferred in the Filtered DM scenario. We also assume the sound speed inside the wall to be $c_{\chi}^{\mr{in}} = \sqrt{T_{\chi,f}^{\mr{out}}/m_{\chi}^{\mr{in}}}$ with  $m_{\chi}^{\mr{in}} / T_{\chi,f}^{\mr{out}} = 30$. In the deflagration branch, the fluid velocity inside the wall is larger than that outside the wall. Therefore, compared to the case where hydrodynamic effects are neglected, the amount of DM that can penetrate into the inside of the wall is reduced. It is evident from the behavior shown in Fig.~\ref{fig: solution for velocities} that hydrodynamic effects become more important as the effective latent heat increases. On the other hand, in the limit of a slow wall velocity, the ratio approaches unity regardless of the value of the effective latent heat. This is because, as shown in Fig.~\ref{fig: solution for velocities}, the deflagration branch intersects at the origin in the $(v_{\chi, f}^{\mr{out}}, v_{\chi, f}^{\mr{in}})$ plane.

Finally, we comment on the lower limit of the fluid velocity from the diffusion perspective. In this section, we estimate the amount of DM entering the wall using an expression that neglects the outflow induced by thermal fluctuations. However, as briefly mentioned in Sec.~\ref{subsec: Filtering Effect}, this approximation is not justified when the wall velocity is extremely small. As a threshold below which the outflow cannot be neglected, Sec.~\ref{subsec: Filtering Effect} adopted 
$v_{\chi}^{\mr{th}} := \sqrt{3 T_n / m_{\chi}^{\mr{in}}}$.
However, from the perspective of fluid dynamics, this lower limit remains unsatisfactory. In fact, the impact of thermal diffusion, which can drive the outflow, differs depending on two distinct regimes:

\begin{itemize}
\item \textbf{LTE regime}: 
When LTE is valid, the mean free path is shorter than the wall thickness, and thermal diffusion can therefore be a source of particle escape. 
The diffusion velocity can be estimated as 
$v^{\mr{diff}} \sim v^{\mr{th}} L_{\mr{MFP}}/L_w$. 
Thus, a rough lower bound would be $v_{\chi,f}^{\mr{out}} \gtrsim v^{\mr{diff}}$. 
Since diffusion is suppressed when the mean free path is short, particle escape can be suppressed even if the wall velocity is slower than $v^{\mr{th}}$. 
However, in Eq.~\eqref{eq: number density inside}, the apparent divergence of the prefactor that appears in the limit $v_w \to 0$ should be regulated by relaxation processes that enforce LTE.
\item \textbf{Ballistic regime}: 
In the ballistic regime, where the mean free path is long, the effect of thermal diffusion near the wall can be neglected. 
However, this does not mean that diffusion can be ignored even in the limit where the wall is at rest. 
If the wall velocity becomes sufficiently small, the mean free path $L_{\mr{MFP}}$, which is determined by the inverse interaction rate and the particle velocity, becomes shorter than the wall thickness. 
In that case, the system transitions to the LTE regime.
\end{itemize}
\section{Summary}
\label{sec: Summary}
In this work, we have studied the hydrodynamic behavior in the setup of a Filtered DM scenario and have investigated how these effects influence the relic abundance of Filtered DM.
In the Filtered DM scenario, the dynamics of the FOPT differ from the situation typically considered for the electroweak FOPT. For DM particles, the filtering effect plays an important role, whereas the bubble wall has a negligible impact on the radiation particles. To properly treat this difference, we formulated the hydrodynamics of this system as a two-component fluid composed of DM and radiation. From the conservation of the energy--momentum tensor across the bubble wall, we derived the matching conditions for the fluid variables on the two sides of the wall. These matching conditions incorporate not only the hydrodynamic obstruction associated with crossing the wall but also effects originating from the DM fluid component that cannot penetrate the wall: the reflection of DM by the wall, and the conversion of energy and momentum from the DM fluid to the radiation fluid through relaxation processes.

Depending on the hierarchy between the mean free path and the wall thickness in the plasma frame, the hydrodynamic behavior during the FOPT can be classified into two regimes. One is the ballistic regime, realized when the mean free path is sufficiently long, and the other is the LTE regime, realized when the mean free path is short~\cite{Ai:2024btx, Krajewski:2024xuz}. DM particles that cannot penetrate the wall are predominantly reflected by the wall in the ballistic regime, whereas in the LTE regime, they are mainly converted into the radiation fluid. Thus, the fate of the DM component that cannot cross the wall differs significantly between the two regimes.

We have parameterized these effects in a manner that is as model-independent as possible and have shown that the solutions can be classified into detonation-like and deflagration-like branches. The difference between the ballistic and LTE regimes affects the hydrodynamics through different parameterizations of the effective latent heat used to characterize the solutions. We found that, in both regimes, the effects originating from the DM component that cannot cross the wall impose stringent conditions for realizing slow deflagration branches.

Based on the obtained hydrodynamic solutions, we revisited the analysis of the impact of hydrodynamic effects on the relic abundance of Filtered DM, as pointed out in Ref.~\cite{Jiang:2023nkj}. We have incorporated effects that were not included in the previous study, namely the contribution from the DM component that cannot penetrate the wall and the difference in the sound speed inside the wall~\cite{Leitao:2014pda, Giese:2020rtr, Giese:2020znk, Ai:2023see, Sanchez-Garitaonandia:2023zqz}. We found that, even after taking these effects into account, the change in the velocity profile across the wall modifies the relic abundance compared to the case where hydrodynamic effects are neglected.

We have also revisited the evolution of the entropy current. In particular, we showed that when there exists a transfer of energy and momentum from the DM fluid to the radiation fluid, the entropy current is not necessarily conserved even if LTE is assumed. Moreover, under the planar approximation, we found that negative entropy production can emerge when the entropy current is integrated in the vicinity of the wall in the LTE regime.

Motivated by this observation, we suggested that the Filtered DM system may be regarded as an information-thermodynamic system, in the sense that the wall effectively measures the particle species and applies feedback based on the outcome.  From this perspective, the conventional second law of thermodynamics for the plasma sector may not necessarily be satisfied, while the second law is restored when the scalar field is included as part of the full system. Further investigation in this direction would be of interest.

\section*{Acknowledgments}
The author would like to express sincere thanks to Kazuki Enomoto for collaboration and many useful discussions at the early stage of this work. The author also acknowledges useful discussions with Wen Yin. This work was supported by JSPS KAKENHI (Grant Numbers 25KJ0022).

\appendix

\section{Derivation of entropy current equation}
\label{app: Derivation of entropy current}

In this Appendix, we present the derivation of the evolution equations for the entropy current, Eqs.\eqref{eq: entropy non-conservation in dark sector} and \eqref{eq: entropy non-conservation in SM sector}, introduced in the main text.

First, we consider the DM fluid sector. Rewriting Eq.~\eqref{eq: energy momentum evolution for dark matter plasma} as the evolution equation for $T_{\chi,f}^{\mu\nu}$, we obtain the following form:
\begin{align}
\label{appeq: energy momentum evolution for dark matter plasma}
\partial_{\mu} T_{\chi, f}^{\mu\nu} 
	&= \partial^{\nu} \phi \, \frac{d m_{\chi}^2}{d\phi} \int \frac{d^3 \bs{p}}{(2\pi)^3} \frac{1}{2 E_{\chi}} (f_{\chi, f} + f_{\chi, r}) 
		+ \Theta_{\chi}^{\nu} - \Xi_{\chi}^{\nu}.
\end{align}

Assuming the perfect fluid form on $T_{\chi,f}$ and redefining the pressure by absorbing the potential contribution as 
$p_{\chi} = p_{\chi,f} - V_0(\phi)$,
we obtain
\begin{align}
\label{appeq: energy momentum evolution for dark matter plasma 2}
\partial_{\mu} T_{\chi, f}^{\mu\nu} 
	&= \partial_{\mu} (\omega_{\chi, f} u_{\chi,f}^{\mu} u_{\chi,f}^{\nu} - p_{\chi} g^{\mu\nu})
	- V'_0(\phi) \partial^{\nu} \phi ,
\end{align}
where we note that  $\omega_{\chi, f} = \rho_{\chi} + p_{\chi} (= \rho_{\chi, f} + p_{\chi, f})$ with $\rho_{\chi} = \rho_{\chi, f} + V_0(\phi)$, and the prime denotes differentiation with respect to the scalar field.

On the other hand, for the right-hand side in Eq.~\eqref{appeq: energy momentum evolution for dark matter plasma}, the contribution from the equilibrium distribution in the first term can be regarded as a correction to the effective potential~\cite{Moore:1995si}, because it also appears in the scalar field equation of motion~\eqref{eq: scalar eom}. Thus,
\begin{align}
\label{eq: scalar eom}
\frac{d m_{\chi}^2}{d\phi} 
	\int \frac{d^3 \bs{p}}{(2\pi)^3} \frac{1}{2 E_{\chi}} f_{\chi}^{\mr{eq}}
= V'_{1}(\phi, T),
\end{align}
where $V'_{1}(\phi, T)$ denotes the one-loop correction to the effective potential.

Therefore, contracting both sides of Eq.~\eqref{appeq: energy momentum evolution for dark matter plasma} with the fluid velocity $(u_{\chi,f})_\nu$ yields
\begin{align}
\label{appeq: energy momentum evolution for dark matter plasma 2}
&(u_{\chi,f})_{\nu} \partial_{\mu} (\omega_{\chi,f} u_{\chi,f}^{\mu} u_{\chi,f}^{\nu} - p_{\chi} g^{\mu\nu})
	- V'_{\mr{th}}(\phi) (u_{\chi,f})_{\nu} \partial^{\nu} \phi \nonumber\\
	&= (u_{\chi,f})_{\nu} \partial^{\nu} \phi \, \frac{d m_{\chi}^2}{d\phi} 
	\int \frac{d^3 \bs{p}}{(2\pi)^3} \frac{1}{2 E_{\chi}} (\delta f_{\chi, f} + f_{\chi, r}) \nonumber\\
    &\quad
	+ (u_{\chi,f})_{\nu} ( \Theta_{\chi}^{\nu} - \Xi_{\chi}^{\nu}).
\end{align}
where $V_{th}(\phi) := V_0(\phi) + V_1 (\phi, T)$ denotes thermally corrected potential.

Here, for derivative  of $p_{\chi}$ we can use\footnote{
We note that, from the Gibbs--Duhem equation,
\begin{align}
d p = s\, dT + n\, d \mu + \left({\partial p \over \partial \phi}\right) d \phi,
\end{align}
where the subscripts representing the DM fluid have been omitted.
}
\begin{align}
\partial_{\mu} p_{\chi} 
	&= s_{\chi, f},\partial_{\mu} T_{\chi,f} 
    + n_{\chi,f}\, \partial_{\mu} \mu_{\chi}
	- V'_{\mr{th}}(\phi)\,\partial_{\mu} \phi.
\end{align}
In addition, since $\omega_{\chi, f} = s_{\chi, f} T_{\chi, f} + \mu_{\chi} n_{\chi, f}$, we obtain
\begin{align}
&(u_{\chi,f})_{\nu}\, \partial_{\mu} (\omega_{\chi,f} u_{\chi,f}^{\mu} u_{\chi,f}^{\nu})\nonumber\\
	&= T_{\chi,f}\, \partial_{\mu} (s_{\chi,f} u_{\chi,f}^{\mu})
	+ \mu_{\chi}\, \partial_{\mu} (n_{\chi,f} u_{\chi,f}^{\mu}) \nonumber\\
    &\quad
        +s_{\chi, f},\partial_{\mu} T_{\chi,f} 
        + n_{\chi,f}\, \partial_{\mu} \mu_{\chi}.
\end{align}

Therefore, substituting the above relations into Eq.~\eqref{appeq: energy momentum evolution for dark matter plasma 2}, we obtain:
\begin{align}
\label{appeq: entropy non-conservation in dark sector}
\partial_{\mu} S_{\chi, f}^{\mu}
	 &=-{\mu_{\chi} \over T_{\chi, f}} \, \partial_{\mu} \bigl(n_{\chi, f} \, u_{\chi, f}^{\mu}\bigr)
	   + {u_{\chi, f}^{\mu} \over T_{\chi,f}} \, \left(\Theta_{\chi, \mu} - \Xi_{\chi, \mu} \right)\nonumber\\
    &\quad
        +{u_{\chi,f}^{\mu} \over T_{\chi,f}}
            (\partial_{\mu} \mu_{\chi})
            \int \frac{d^3 p}{(2\pi)^3} {1 \over E_{\chi}} (\delta f_{\chi, f} + f_{\chi, r}) .
\end{align}
We note that in Eq.~\eqref{eq: entropy non-conservation in dark sector} of the main text, $\delta f_{\chi,f}$ is treated as a small perturbation and is therefore neglected.

The evolution equation for the entropy current on the radiation side can be derived in the same manner. The only difference from the DM fluid case is that there is no need to take a reflected component into account.

\section{Matching equations without reflection or transfer}
\label{app: Matching equations without reflection or transfer}
In this Appendix, we review how to solve the hydrodynamic matching conditions without reflection or transfer effect. Even when the reflection or transfer effects are included, as discussed in the main text, an appropriate reparametrization reduces these equations to the matching conditions without reflection or transfer effects. 

Let us begin with the explicit form of the matching conditions to be solved, which are derived from energy-momentum conservation~\cite{Steinhardt:1981ct, Kurki-Suonio:1984zeb, Landau:1989, Laine:1993ey, Ignatius:1993qn}:
\begin{align}
\label{appeq: energy flux conservation}
&\omega^{\mr{out}} (\gamma^{\mr{out}})^2 v^{\mr{out}} 
- \omega^{\mr{in}} (\gamma^{\mr{in}})^2 v^{\mr{in}} 
= 0 \\
\label{appeq: momentum conservation}
&\omega^{\mr{out}} (\gamma^{\mr{out}})^2 (v^{\mr{out}})^2 
- \omega^{\mr{in}} (\gamma^{\mr{in}})^2 (v^{\mr{in}})^2 \nonumber \\
&\hspace{5em}
= p^{\mr{in}} - p^{\mr{out}} + \Delta \epsilon,
\end{align}
where we have omitted labels introduced to distinguish DM and radiation, because they are no longer meaningful.

Following Refs.~\cite{Steinhardt:1981ct, Kurki-Suonio:1984zeb, Landau:1989, Laine:1993ey, Ignatius:1993qn}, we first solve these equations for the ratio $v^{\mr{out}}/v^{\mr{in}}$ and the product $v^{\mr{out}} v^{\mr{in}}$ of the fluid velocities outside and inside the wall. To make it clear that this derivation does not depend on the equation of state, namely the sound speed of the system, we explicitly present the calculation step by step.

If we introduce the energy flux $J$ as follows, Eq.~\eqref{appeq: energy flux conservation} shows that it is a conserved quantity:
\begin{align}
J := \omega^{\mr{out}} (\gamma^{\mr{out}})^2 v^{\mr{out}} (= \omega^{\mr{in}} (\gamma^{\mr{in}})^2 v^{\mr{in}}),
\end{align}
Then equation~\eqref{appeq: momentum conservation} can be rewritten in terms of $J$ as
\begin{align}
J (v^{\mr{out}} - v^{\mr{in}}) = p^{\mr{in}} - p^{\mr{out}}.
\end{align}
Substituting
\begin{align}
J v^{\mr{out}} &= \omega^{\mr{in}} (\gamma^{\mr{in}})^2 v^{\mr{in}} v^{\mr{out}}, \\
J v^{\mr{in}} &= \omega^{\mr{out}} (\gamma^{\mr{out}})^2 v^{\mr{in}} v^{\mr{out}},
\end{align}
into this equation, we obtain
\begin{align}
v^{\mr{out}} v^{\mr{in}}  \Big(\omega^{\mr{in}} (\gamma^{\mr{in}})^2 - \omega^{\mr{out}} (\gamma^{\mr{out}})^2\Big)
= p^{\mr{in}} - p^{\mr{out}}.
\end{align}

The expression inside the parentheses is
\begin{align}
&\omega^{\mr{out}} (\gamma^{\mr{out}})^2 - \omega^{\mr{in}}(\gamma^{\mr{in}})^2 \nonumber\\
    &= \omega^{\mr{out}} - \omega^{\mr{in}}
        + \left(\omega^{\mr{out}} (v^{\mr{out}})^2 (\gamma^{\mr{out}})^2
            - \omega^{\mr{in}} (v^{\mr{in}})^2 (\gamma^{\mr{in}})^2 \right)\nonumber\\
    &= \rho^{\mr{out}} - \rho^{\mr{in}} .
\end{align}
Here, in the second equality, we used Eq.~\eqref{appeq: momentum conservation} and
$\omega^{\mr{out}(\mr{in})} = \rho^{\mr{out}(\mr{in})} + p^{\mr{out}(\mr{in})}$. Therefore, for the product of the velocities, we obtain
\begin{align}
\label{appeq: solution for product of the velocities}
v^{\mr{out}} v^{\mr{in}}
	= \frac{p^{\mr{in}} - p^{\mr{out}}}{\rho^{\mr{out}} - \rho^{\mr{in}}}.
\end{align}

On the other hand, for Eq.~\eqref{appeq: energy flux conservation}, by substituting the definition of the gamma factor,
$\gamma^{\mr{out}(\mr{in})} = 1/(1-(v^{\mr{out}(\mr{in})})^2)$,
and the definition of the enthalpy,
$\omega^{\mr{out}(\mr{in})} = \rho^{\mr{out}(\mr{in})} + p^{\mr{out}(\mr{in})}$, we obtain
\begin{align}
v^{\mr{out}} (\rho^{\mr{out}} + p^{\mr{out}}) (1- (v^{\mr{in}})^2)
	= v^{\mr{in}} (\rho^{\mr{in}} + p^{\mr{in}}) (1- (v^{\mr{out}})^2) .
\end{align}
By straightforward algebraic manipulation, this can be reduced to
\begin{align}
{v^{\mr{out}} \over v^{\mr{in}}}
	= \frac{\rho^{\mr{in}} + p^{\mr{out}} + v^{\mr{out}} v^{\mr{in}} (\rho^{\mr{out}} + p^{\mr{out}})}{\rho^{\mr{out}}+  p^{\mr{out}} + v^{\mr{out}} v^{\mr{in}} (\rho^{\mr{in}} + p^{\mr{in}})} .
\end{align}

Substituting Eq.~\eqref{appeq: solution for product of the velocities} into the right-hand side of this expression and again performing straightforward algebraic manipulation, we obtain
\begin{align}
\label{appeq: solution for ratio of the velocities}
{v^{\mr{out}} \over v^{\mr{in}}}
	= \frac{\rho^{\mr{in}} + p^{\mr{out}}}{\rho^{\mr{out}} +p^{\mr{in}}}.
\end{align}

The subsequent calculation depends on the equation of state. Following Ref.~\cite{Giese:2020rtr}, we decompose the pressure difference as
\begin{align}
\Delta p
	 &:= p^{+}(T^{\mr{out}}) - p^{-}(T^{\mr{in}}) \nonumber \\
	 &= \left(p^{+}(T^{\mr{out}})
     - p^{-}(T^{\mr{out}})\right)\nonumber\\
     &\quad
			+ \left(p^{-}(T^{\mr{out}}) - p^{-}(T^{\mr{in}})\right)\\
	&=: D p + \delta p,
\end{align}
and an analogous definition is adopted for the difference in the energy density, denoted as $\Delta \rho$.

If the FOPT is not too strong, one has $T^{\mr{out}} \simeq T^{\mr{in}}$, and therefore~\cite{Giese:2020rtr}
\begin{align}
{\delta p\over \delta \rho} \simeq (c_s^{\mr{in}})^2,
\end{align}
where $c_s^{\mr{in}}$ denotes the sound speed inside the wall.

Rewriting Eq.~\eqref{appeq: solution for product of the velocities} using this sound speed yields~\cite{Giese:2020rtr}
\begin{align}
\delta p \left(1 - \frac{v^{\mr{out}} v^{\mr{in}}}{(c_{s}^{\mr{in}})^2}\right)
	= v^{\mr{out}} v^{\mr{in}} D \rho - Dp .
\end{align}

Using this relation, $\Delta p$ and $\Delta \rho$ can be expressed in terms of the sound speed as
\begin{align}
\Delta p
	&= \frac{v^{\mr{out}} v^{\mr{in}} \left(D \rho- Dp/ (c_{s}^{\mr{in}})^2 \right)}{(1-v^{\mr{out}} v^{\mr{in}})/ (c_{s}^{\mr{in}})^2}, \\
\Delta \rho
	&= \frac{D \rho- Dp/ (c_{s}^{\mr{in}})^2}{(1-v^{\mr{out}} v^{\mr{in}})/ (c_{s}^{\mr{in}})^2}.
\end{align}
On the other hand, since $\omega^{\mr{out}(\mr{in})} = \rho^{\mr{out}(\mr{in})} + p^{\mr{out}(\mr{in})}$, Eq.~\eqref{appeq: solution for ratio of the velocities} can be rewritten as
\begin{align}
\frac{v^{\mr{out}}}{v^{\mr{in}}}
	=\frac{1 - \Delta \rho /\omega^{\mr{out}}}{1- \Delta p/ \omega^{\mr{out}}}.
\end{align}
Substituting $\Delta p$ and $\Delta \rho$ into this expression, we obtain~\cite{Giese:2020rtr, Giese:2020znk}
\begin{align}
\label{appeq: ratio relation 2}
\frac{v^{\mr{out}}}{v^{\mr{in}}}
	= \frac{\left(v^{\mr{out}} v^{\mr{in}}/(c_s^{\mr{in}})^2 -1\right) + 3 \alpha_{\bar{\theta}}}
			{\left(v^{\mr{out}} v^{\mr{in}}/(c_s^{\mr{in}})^2 -1\right) + 3 v^{\mr{out}}v^{\mr{in}} \alpha_{\bar{\theta}}},
\end{align}
where we have defined the following parameter\footnote{
In models where the sound speed inside the wall is strictly constant (the so-called template model or $\nu$ model)~\cite{Leitao:2014pda, Giese:2020rtr, Giese:2020znk}, this parameter is related to the dimensionless latent heat $\alpha^{\mr{out}} = 3 \Delta \epsilon / 4 \omega^{\mr{out}}$ as
\begin{align}
\alpha_{\bar{\theta}}
	= \frac{1}{12 (c_{s}^{\mr{in}})^2}
	\left[
	(3 c_{s}^{\mr{in}})^2 - 1
	+ 3 \alpha^{\mr{out}} \left(1 + (c_{s}^{\mr{in}})^2\right)
	\right].
\end{align}
}
\begin{align}
\alpha_{\bar{\theta}}
	:= \frac{D\bar{\theta}}{3 \omega^{\mr{out}}},
\quad
\bar{\theta} 
	= \rho - \frac{p}{(c_{s}^{\mr{in}})^2}.
\end{align}

Finally, we solve Eq.~\eqref{appeq: ratio relation 2} for the velocity inside the wall, $v^{\mr{in}}$. Introducing $y = v^{\mr{in}}/v^{\mr{out}}$, we have $v^{\mr{in}} v^{\mr{out}} = y (v^{\mr{out}})^2$, and thus Eq.~\eqref{appeq: ratio relation 2} can be rewritten as
\begin{align}
y
	= \frac{\left(y (v^{\mr{out}})^2 - (c_{s}^{\mr{in}})^2\right) + 3 y (v^{\mr{out}})^2 (c_{s}^{\mr{in}})^2\alpha_{\bar{\theta}}}
			{\left(y (v^{\mr{out}})^2 - (c_{s}^{\mr{in}})^2\right) + 3 (c_{s}^{\mr{in}})^2 \alpha_{\bar{\theta}}}.
\end{align}
Alternatively, assuming that the denominator is nonzero, this is equivalent to the quadratic equation for $y$,
\begin{align}
&(v^{\mr{out}})^2 y^2
-\left[(c_{s}^{\mr{in}})^2 (1 + 3 \alpha_{\bar{\theta}})
+ (v^{\mr{out}})^2 (1 + 3 \alpha_{\bar{\theta}})\right] y \nonumber\\
    &\hspace{10em}
            + (c_{s}^{\mr{in}})^2
            = 0.
\end{align}

Solving this equation for $y$ and then multiplying by $v^{\mr{out}}$ to recover $v^{\mr{in}}$, we obtain
\begin{align}
\label{appeq: solution of velocity inside}
v^{\mr{in}}
	&= \frac{1}{2 v^{\mr{out}}}
        \Bigg[
            G(v^{\mr{out}}, \alpha_{\bar{\theta}})\nonumber \\
        &\hspace{1em}
            \pm \sqrt{ G(v^{\mr{out}}, \alpha_{\bar{\theta}})^2 - 4 (v^{\mr{out}})^2 (c_{s}^{\mr{in}})^2}
            \Bigg],
\end{align}
where $G(v, \alpha)$ is the same function as that defined in Eq.~\eqref{eq: G definition} in the main text.

\begin{figure*}[t]
  \centering
  \begin{minipage}[t]{0.48\textwidth}
    \centering
    \includegraphics[width=\textwidth]{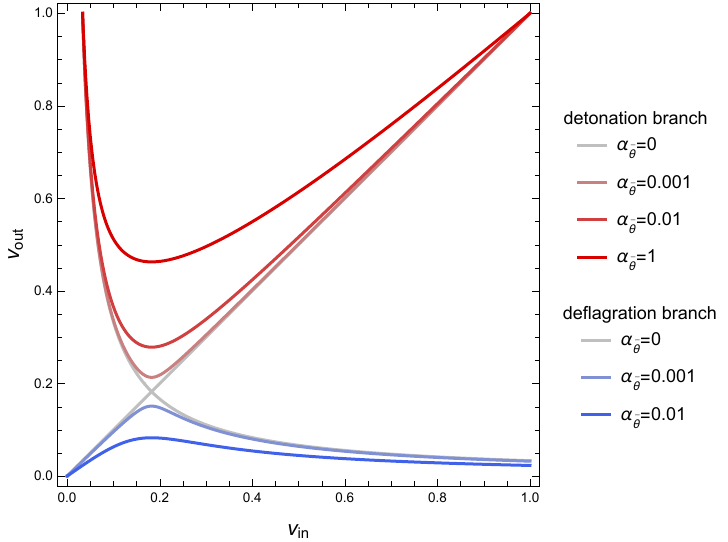}

    \vspace{0.5em}
    \small (left) Non-relativistic sound speed $c_{\chi}^{\mr{in}} = \sqrt{T_{\chi,f}^{\mr{out}}/m_{\chi}^{\mr{in}}}$ with the benchmark choice $m_{\chi}^{\mr{in}} / T_{\chi,f}^{\mr{out}} = 30$
  \end{minipage}
  \hfill
  \begin{minipage}[t]{0.505\textwidth}
    \centering
    \includegraphics[width=\textwidth]{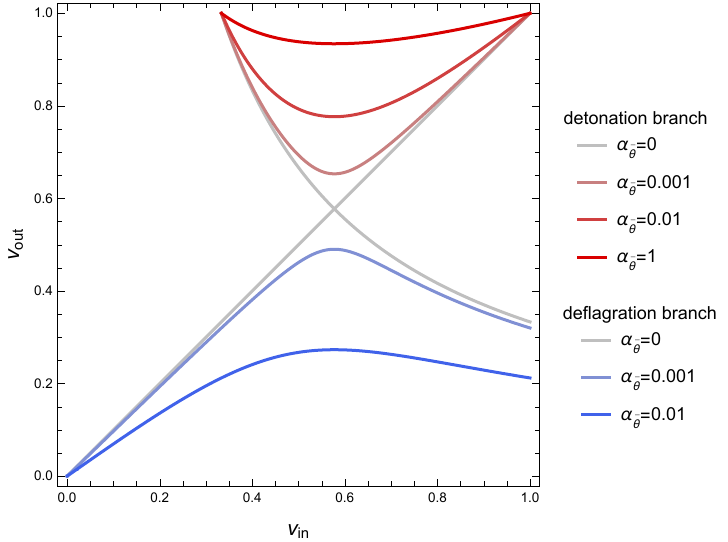}

    \vspace{0.5em}
    \small (right) Radiation sound speed $c_{\chi}^{\mr{in}} = 1/\sqrt{3}$
  \end{minipage}

  \caption{The relation between the DM fluid velocities inside and outside the wall obtained from the matching conditions without reflection or transfer effects. The benchmark values of $\alpha_{\bar{\theta}}$ are the same as those shown in Fig.~\ref{fig: solution for velocities} in the main text for the detonation branch, while different values are used for the deflagration case in the figure.  In both panels, the realizable velocity configurations are restricted to the cases where $(v_{\chi,f}^{\mr{out}}, v_{\chi,f}^{\mr{in}})$ are either both supersonic or both subsonic~\cite{Steinhardt:1981ct, Landau:1989, Laine:1993ey, Leitao:2014pda}.
  }
  \label{fig: solution for velocities wo}
\end{figure*}

Alternatively, although it is not used in the main text, solving for $v^{\mr{out}}$ yields
\begin{align}
\label{appeq: solution of velocity outside}
v^{\mr{out}}
	&=
	\frac{
	F(v^{\mr{in}}) \pm \sqrt{F(v^{\mr{in}})^2 + (c_{s}^{\mr{in}})^2 (1- 3 \alpha_{\bar{\theta}}) \left(1 + 3 (c_{s}^{\mr{in}})^2 \alpha_{\bar{\theta}}\right)}
	}{
		v^{\mr{in}} \left(1 + 3 (c_{s}^{\mr{in}})^2 \alpha_{\bar{\theta}}\right)
	}.
\end{align}
Here $F(v)$ is defined as
\begin{align}
F(v) := \frac{v}{2} + \frac{c_{s}^{\mr{in}}}{2}.
\end{align}
In Fig.~\ref{fig: solution for velocities wo}, we present the solutions of Eq.~\eqref{appeq: solution of velocity inside}. For the benchmark values of the sound speed, we adopt the same choices as in the main text: a non-relativistic gas (left panel) and a relativistic gas (right panel). Similarly, for the DM mass and the temperature ratio, we take $m_{\chi}^{\mr{in}} / T_{\chi,f}^{\mr{out}} = 30$ as the benchmark value.

\bibliographystyle{utphysmod}
\bibliography{ref}

\end{document}